\documentclass[10pt,a4paper]{article}

\usepackage[english]{babel}
\usepackage{comment}

\usepackage{amsmath}
\usepackage{amssymb}
\usepackage{graphicx,epsfig}
\usepackage[mathscr]{eucal}

\usepackage{color}
\usepackage[numbers,sort&compress]{natbib}
\usepackage{multirow}
\usepackage{tabularray}
\usepackage{tabularx}
\UseTblrLibrary{booktabs}
\usepackage{float}

\usepackage{slashed}
\usepackage{braket}

\usepackage{scalefnt,ulem,pstricks}

\usepackage{xspace}
\usepackage{setspace}
\usepackage{xstring}
\usepackage{mathrsfs}
\usepackage{amsbsy}

\newcommand{\MSbar}{\overline{\mathrm{MS} }}
\newcommand{\alphaEW}{\alpha^{[h]}(\mu)}
\newcommand{\alphaEWnl}{\alpha(\mu)}

\newcommand{\fin}{\mathrm{fin}}
\newcommand{\sing}{\mathrm{sing}}

\newcommand{\produc}{\mathrm{prod}}
\newcommand{\dec}{\mathrm{dec}}

\newcommand{\M}{\mathcal{M}}
\newcommand{\Z}{\mathcal{Z}}

\newcommand{\deltaren}{\tilde{\delta}}

\newcommand{\Mren}{\widetilde{\mathcal{M}}}
\newcommand{\Mrendc}{{\mathcal{\hat{M}}}}

\newcommand{\ONp}{\tilde{p}}
\newcommand{\OFFp}{p}

\newcommand{\PA}{\mathrm{PA}}
\newcommand{\fact}{\mathrm{fact}}
\newcommand{\sm}{\scalebox{0.4}[0.8]{$-$}}
\newcommand{\nf}{\mathrm{non\sm fact}}
\newcommand{\nfbis}{\mathrm{nf}}

\newcommand{\D}{\mathcal{D}}
\newcommand{\C}{\mathcal{C}}
\newcommand{\B}{\mathcal{B}}
\newcommand{\DD}{\mathcal{D\hspace{-0.05cm}D}}
\newcommand{\DC}{\mathcal{D\hspace{-0.05cm}C}}
\newcommand{\DB}{\mathcal{D\hspace{-0.05cm}B}}
\newcommand{\CC}{\mathcal{C\hspace{-0.03cm}C}}
\newcommand{\CB}{\mathcal{C\hspace{-0.05cm}B}}
\newcommand{\BB}{\mathcal{B\hspace{-0.05cm}B}}

\newcommand{\musoft}{1/\bar{t}}
\newcommand{\musoftsq}{1/\bar{t}^2}

\usepackage{scalefnt,pstricks}
\usepackage{cancel}

\newcommand{\refeq}[1]{Eq.~(\ref{#1})}
\newcommand{\refeqs}[1]{Eqs.~(\ref{#1})}

\usepackage{array}
\newcolumntype{L}[1]{>{\raggedright\let\newline\\\arraybackslash\hspace{0pt}}m{#1}}
\newcolumntype{C}[1]{>{\centering\let\newline\\\arraybackslash\hspace{0pt}}m{#1}}
\newcolumntype{R}[1]{>{\raggedleft\let\newline\\\arraybackslash\hspace{0pt}}m{#1}}

\usepackage[font=normal, labelfont=bf]{caption}

\usepackage{lipsum}
\usepackage[colorlinks=false,filecolor=black,urlcolor=cyan]{hyperref}

\begin{document}
\begin{titlepage}
\begin{flushright}
ZU-TH 16/26 \\
TUM-HEP-1603/26 \\
CERN-TH-2026-103
\end{flushright}

\renewcommand*{\thefootnote}{\fnsymbol{footnote}}
\vspace*{0.5cm}

\begin{center}
  {\Large \bf Non-factorisable electroweak virtual corrections \\[0.25cm]
    to single-resonant processes}
\end{center}

\par \vspace{2mm}
\begin{center}
    {\bf Fazila Ahmadova${}^{(a)}$}, {\bf Luca Buonocore${}^{(b)}$}, \\[0.25cm]
    {\bf Massimiliano Grazzini${}^{(a)}$} and {\bf Chiara Savoini${}^{(c)}$}

\vspace{5mm}
${}^{(a)}$Physik Institut, Universit\"at Z\"urich, 8057 Z\"urich, Switzerland\\[0.25cm]

${}^{(b)}$Theoretical Physics Department, CERN, CH-1211 Geneva 23, Switzerland\\[0.25cm]

${}^{(c)}$Technical University of Munich, TUM School of Natural Sciences, Physics Department, James-Franck-Stra{\ss}e 1, 85748 Garching, Germany

\vspace{5mm}

\end{center}

\par \vspace{2mm}
\begin{center} {\large \bf Abstract}

\end{center}
\begin{quote}
\pretolerance 10000

We consider electroweak (EW) virtual corrections to $2\to 2$ fermion scattering processes mediated by a vector boson $V$ ($V=W^\pm,Z$) in the pole approximation.
As is well known, the computation can be organised into factorisable and non-factorisable contributions.
The factorisable corrections can be computed by evaluating the (polarised) EW form factor of the vector boson at the relevant perturbative order. The non-factorisable corrections are instead driven by soft-photon exchanges between the initial- and final-state fermions and/or the resonance. We perform an explicit two-loop computation to show that, once the heavy degrees of freedom are properly decoupled, such non-factorisable corrections can be expressed as an iteration of the one-loop result, plus a new contribution due to (light) fermion loops. The final two-loop result, which can be expected on general grounds from soft-photon factorisation, is shown to hold exactly in dimensional regularisation and is peculiar to the exchange of a single resonance. We discuss its extension to all perturbative orders.

\end{quote}

\vspace*{\fill}
May 2026
\end{titlepage}

\tableofcontents

\renewcommand{\thefootnote}{\fnsymbol{footnote}}

\section{Introduction}
\label{sec:intro}

More than fifteen years after the start of the Large Hadron Collider (LHC) physics programme, the Standard Model (SM) remains, to the best of our knowledge, the theory that describes elementary particles and their interactions. 
Despite some intriguing hints of possible new-physics signals, no conclusive evidence for new particles or phenomena has been observed, and the scalar resonance discovered in 2012 is fully consistent with the SM Higgs boson. This suggests that physics beyond the SM, although expected on general grounds, may lie at energy scales not directly accessible in the near future.

The forthcoming High-Luminosity upgrade of the LHC will provide a huge sample of high-quality data. 
In this context, new physics may manifest itself through subtle deviations from SM predictions, for instance, in the tails of kinematic distributions or as small excesses over expected backgrounds. This makes precise theoretical calculations essential for a more thorough scrutiny of the Higgs sector and for achieving a reliable control over SM background processes.
In addition, the European Strategy for Particle Physics recommends the electron-positron Future Circular Collider (FCC-ee) as the preferred option for CERN's next flagship collider \cite{deBlas:2025gyz}. Designed to operate at progressively increasing centre-of-mass energies--- starting at the $Z$ pole, then at the $WW$ threshold, followed by operation as a Higgs factory, and ultimately reaching the $t{\bar t}$ threshold---the FCC-ee will enable unprecedented precision in measurements of key SM parameters and observables.
Fully exploiting the potential of these measurements will require substantial improvements in theoretical predictions to match the expected experimental accuracy, necessitating the inclusion of higher-order radiative corrections in both the QCD and electroweak (EW) sectors of the SM. 

Many of the most interesting high-multiplicity processes in the SM involve heavy unstable particles such as the $W$ and $Z$ bosons and the top quark. 
The properties of $W$ and $Z$ bosons have a significant impact on key EW precision observables, being the mediators of the weak interactions, while the top quark contributes to backgrounds in nearly every LHC analysis. 
These resonant states, however, do not appear directly as stable particles in the final state, but must be reconstructed from their decay products, typically energetic jets and leptons accompanied by missing transverse energy.
The need to provide precise theoretical predictions naturally raises the question of how to correctly treat the decay of these intermediate unstable particles.

A complete description requires a fully off-shell calculation, i.e.\ the gauge-invariant set of all diagrams contributing to a given final state. Such a treatment consistently incorporates resonant and non-resonant contributions, as well as finite-width effects. 
However, the technical complexity of the corresponding multiloop amplitudes currently restricts such calculations to next-to-leading order (NLO). Despite continuous progress in precision QCD and EW computations (see e.g.\ Ref.~\cite{Huss:2025nlt} and references therein), the evaluation of multiloop amplitudes with several external legs and/or multiple mass scales remains out of reach.
This limitation is particularly relevant for processes involving heavy, unstable particles, where large final-state multiplicities, increasingly intricate topologies, and multiple scales pose severe challenges. As a result, exact computations are not presently achievable. 
Extending theoretical predictions for such processes to next-to-next-to-leading order (NNLO), therefore, necessitates the use of approximations that exploit the underlying resonant structure.

The simplest approach is the narrow-width approximation (NWA), which models the production and decay of unstable particles while treating them as strictly on-shell.
In this framework, radiative corrections factorise into independent contributions for the production and decay subprocesses, and spin correlations are preserved. 
Off-shell, non-resonant, and non-factorisable effects are completely neglected. 
This approximation is expected to provide reliable predictions for sufficiently inclusive observables, with an accuracy of
${\cal O}(\Gamma/M)$~\cite{Melnikov:1993np,Fadin:1993kt}, where $M$ and $\Gamma$ are the mass and width of the resonance, respectively.
However, its validity becomes questionable in phase-space regions that are particularly sensitive to off-shell effects.
Even when off-shell contributions are negligible for inclusive observables, their impact on more exclusive measurements can be significant and cannot be assumed a priori, motivating the need for approaches that systematically account for finite-width effects.

As is well known, the {\it pole approximation} (PA) provides a consistent way to go beyond the NWA by performing a systematic expansion in the virtuality of the resonant particles (see Ref.~\cite{Denner:2019vbn} and references therein). Within the PA, one distinguishes two classes of gauge-invariant contributions.
The {\it factorisable} corrections comprise all resonant contributions that can be written as a product of on-shell matrix elements for the production and decay subprocesses, which are only connected through the kinematics and spin of the intermediate resonance(s). Their evaluation requires the knowledge of polarised on-shell amplitudes for the production and decay of the resonance(s), which are typically simpler to compute since they involve fewer external legs and mass scales than the full off-shell process.
The {\it non-factorisable} corrections include all remaining resonant contributions that arise from interferences between emissions at different stages of the process, i.e.\ between production and decay. They are driven by the exchange of {\it soft} massless quanta (photons and/or gluons) between the initial- and final-state charged particles and/or the resonance(s)~\cite{Dittmaier:2015bfe}. Although their evaluation requires the computation of specific multileg loop integrals, their soft nature suggests a certain degree of universality.
Overall, compared to a complete off-shell calculation, the PA drastically reduces the computational complexity by focusing on maximally resonant topologies while---unlike the NWA---systematically retaining the leading off-shell and non-factorisable effects in a fully differential manner.

In the past, the PA has been successfully applied to the evaluation of both QCD and EW corrections for several processes.
The first studies in the context of NLO EW corrections were on single resonant $W$ boson production~\cite{Wackeroth:1996hz,Baur:1998kt}.
These early applications were followed by the computation of NLO EW corrections for $W$-pair production with leptonic decays in $e^+e^-$ collisions~\cite{Beenakker:1998gr,Denner:2000bj}, at a time when the corresponding exact one-loop amplitudes were not yet available~\cite{Denner:2005es,Denner:2005fg}.
More recently, the PA has been applied to single top~\cite{Falgari:2010sf,Falgari:2011qa} and top-quark pair hadroproduction~\cite{Falgari:2013gwa} in NLO QCD, and to four-fermion production in $pp$ collisions at NLO EW~\cite{Billoni:2013aba}.
It has also been widely used in the context of polarised vector-boson pair production (see Ref.~\cite{Denner:2025xdz} and references therein).
Comparisons between results obtained in the PA and full off-shell calculations have been carried out for NLO EW corrections to $pp\to W^+W^-\to 4$ lepton production \cite{Biedermann:2016guo}, off-shell $t{\bar t}$ production with leptonic decays~\cite{Denner:2016jyo} and vector-boson scattering~\cite{Biedermann:2016yds}, for NLO QCD corrections to off-shell $t\bar t W^+$ production~\cite{Denner:2020hgg} and for NLO QCD and EW corrections to triple-$W$ production with leptonic decays~\cite{Dittmaier:2019twg} at the LHC.
These studies show remarkable agreement for both inclusive cross sections and differential distributions that are insensitive to non-resonant effects. 

The PA has also been successfully applied to the evaluation of mixed NNLO QCD-EW corrections to the Drell-Yan process~\cite{Dittmaier:2015rxo,Buonocore:2021rxx,Dittmaier:2024row}. In this case, the non-factorisable corrections exhibit a relatively simple structure~\cite{Dittmaier:2014qza}.
Finally, the PA has proven essential for the approximate NNLO QCD computation of $pp\to t\bar t \to W^+W^-b\bar b$~\cite{Buonocore:2025fqs}, where the main missing ingredient is represented by the two-loop non-factorisable corrections. 
Their evaluation poses a major technical challenge and currently prevents extending the PA to full NNLO accuracy.
It is worth emphasising that the non-factorisable corrections, although often regarded as subleading due to their small impact on inclusive observables~\cite{Melnikov:1993np,Fadin:1993kt}, are by no means negligible. As shown in Ref.~\cite{Buonocore:2025fqs}, their impact on fiducial cross sections is small only when real and virtual non-factorisable contributions are consistently removed. At the differential level, however, they can induce sizeable distortions in kinematic distributions, making their inclusion mandatory for precision phenomenology.

In this paper we investigate the structure of non-factorisable EW virtual corrections at two-loop order in single-resonance processes.
By expressing these contributions in terms of suitable soft integrals, we compute them explicitly and show that, after properly decoupling heavy degrees of freedom, the two-loop result can be written in dimensional regularisation as an iteration of the corresponding one-loop contribution, supplemented by a genuinely new term induced by light-fermion loops.
Building on this result, we propose a factorisation formula that extends this structure to all orders in perturbation theory. 
When combined with the factorisable corrections, which require the two-loop polarised EW form factor, our findings provide the basis for constructing approximate two-loop EW amplitudes for key processes such as Drell-Yan lepton-pair production and $e^+e^- \to \mu^+\mu^-$.
More generally, this work constitutes a first step towards the treatment of non-factorisable corrections at higher perturbative orders in more complex scenarios, including processes with multiple resonances and extensions to QCD.

The paper is organised as follows. 
In Section~\ref{sec:calculation} we present our calculation. After outlining the general setup in Section~\ref{sec:preliminaries}, we review the one-loop results in Section~\ref{sec:oneloop}. The new two-loop computation is described in Section~\ref{sec:twoloop}. We begin by analysing contributions from diagrams with two soft photons exchanged between the production and decay subprocesses (Section~\ref{sec:two_soft-photons}), and then turn to the case of a single soft-photon exchange. In this context, fermionic corrections are discussed first (Section~\ref{sec:fermionic}), followed by the non-abelian contributions (Section~\ref{sec:bosonic}).
The complete results are presented in Section~\ref{sec:results_two-loop-nonfact}. Ultraviolet (UV) renormalisation is addressed in Section~\ref{sec:results_UVren}, while Section~\ref{sec:results_IRpoles} is devoted to the analysis of the infrared (IR) structure and the construction of the finite remainder. In Section~\ref{sec:discussion} we discuss our findings and propose an all-order resummation formula for the virtual amplitude in the pole approximation, consistent with our fixed-order results. Finally, our conclusions are summarised in Section~\ref{sec:conclusions}. Further details are provided in two Appendices.

\section{The calculation}
\label{sec:calculation}

\subsection{Preliminaries}\label{sec:preliminaries}
We consider a general four-fermion scattering process
\begin{equation}
	f_1(\OFFp_1)+{\bar f}_2(\OFFp_2)\to V \to f_3(\OFFp_3)+{\bar f}_4(\OFFp_4)  \label{eq:process}
\end{equation}
mediated by the resonance $V$ ($V=W^\pm, Z$). All external fermions are treated as massless.
The corresponding matrix element admits a perturbative expansion in the dimensionless bare coupling $\alpha_0$,
\begin{align}
    \M = (4\pi\alpha_0) \left[
        \M^{(0)} + \left(\frac{\alpha_0}{4\pi}\right)\M^{(1)}+ \left(\frac{\alpha_0}{4\pi}\right)^2\M^{(2)}+\mathcal{O}(\alpha_0^3)
    \right] \,,
    \label{eq:Mbare}
\end{align}
where $\M^{(n)}$ denotes the $n$-loop contribution.
In this work, we focus on the two-loop EW virtual corrections, $\M^{(2)}$, within the pole approximation, which is well justified for observables dominated by the resonant region. 
This framework substantially reduces the complexity of a full off-shell calculation~\footnote{Dedicated efforts are currently underway~\cite{Armadillo:2025mfx,Freitas:2025vax}, but no complete result is yet available.} and offers a systematic path towards the desired result.

For single-resonant processes like in \refeq{eq:process}, the matrix element in~\refeq{eq:Mbare} can be decomposed as
\begin{align}
    \M = \frac{\mathcal{R}(\OFFp_V^2)}{\OFFp_V^2-m_V^2 + \Sigma(\OFFp_V^2)} + \mathcal{N}(\OFFp_V^2) \,,
    \label{eq:PoleExp}
\end{align}
where $\mathcal{R}$ and $\mathcal{N}$ control the resonant and non-resonant parts of the scattering amplitude, respectively, $\Sigma$ denotes the self-energy of the vector boson $V$ and $\OFFp_V$ its off-shell momentum.
For an unstable particle, $\Sigma$ develops a non-vanishing imaginary part and, hence, the propagator acquires a pole in the complex $\OFFp_V^2$ plane at the gauge-invariant location $\mu_V^2 = m_V^2 - i\Gamma_V m_V$, where $m_V$ and $\Gamma_V$ denote the pole mass and width of the vector boson $V$, respectively.
In the PA, the non-resonant term $\mathcal{N}$ in \refeq{eq:PoleExp} is
neglected, while the resonant contribution is asymptotically expanded around
$\OFFp_V^2 = \mu_V^2$ and only the leading term is retained. This contribution
can be organised into two classes of gauge-invariant corrections, namely {\it factorisable} and {\it non-factorisable}.
It is convenient to work in the on-shell (OS) renormalisation scheme for the resonance.
In this scheme, close to the resonance, the propagator in \refeq{eq:PoleExp} can be written as $1/(p_{V}^{2}-\mu_{V}^{2})$ to all orders in perturbation theory, while the residue $\mathcal{R}$ is expanded in the on-shell limit.

At leading order in the EW coupling, only the factorisable contribution is present, and the scattering amplitude $\M^{(0)}$ in \refeq{eq:Mbare} is thus approximated by
\begin{align}
    \M^{(0)}_{\PA} 
    =\frac{1}{\OFFp_V^2 -\mu_V^2} \sum_{\{\vec{h}_I\}}\sum_{\{\vec{h}_F\}}\sum_{{\lambda_V}}
    \M^{(0)}_{\produc}(\vec{h}_I,\lambda_V)\M^{(0)}_{\dec}(\vec{h}_F,\lambda_V) \,,
    \label{eq:M0PA}
\end{align}
where the sums run over the helicity configurations $\vec{h}_I = \{h_1,h_2\}$ and $\vec{h}_F = \{h_3,h_4\}$ of the initial- and final-state particles, and over the intermediate helicity $\lambda_V$ of the resonance.
Here, $\M^{(0)}_{\produc}$ and $\M^{(0)}_{\dec}$ refer to the tree-level on-shell scattering amplitudes for the production and decay subprocesses
\begin{align}
f_1(\ONp_1)+{\bar f}_2(\ONp_2)\to V (\ONp_V)\,, \qquad\qquad\qquad V (\ONp_V)\to f_3(\ONp_3)+{\bar f}_4(\ONp_4) \,, \label{eq:subprocess}
\end{align}
respectively. 
To ensure momentum conservation and on-shell conditions, the external momenta of the fermions have to be mapped $\{\OFFp_n\}\to \{\ONp_n\}$, and the resonance $V$ is evaluated on its mass-shell $\ONp_{V}^2 = m_V^2$.

At higher perturbative orders, radiative corrections affect the production and decay subprocesses independently.
As a result, the factorisable contribution to the $n$-loop virtual amplitude, $\M^{(n)}$, can be constructed as a generalisation of \refeq{eq:M0PA}
\begin{align}
    \M^{(n)}_{\fact} 
    =\frac{1}{\OFFp_V^2 -\mu_V^2} \sum_{m=0}^{n}\sum_{\{\vec{h}_I\}}\sum_{\{\vec{h}_F\}}\sum_{{\lambda_V}}
    \M^{(m)}_{\produc}(\vec{h}_I,\lambda_V)\M^{(n-m)}_{\dec}(\vec{h}_F,\lambda_V) \,, 
    \label{eq:factLloop}
\end{align}
where spin correlations and finite-width effects from the resonance propagator are accounted for.
The polarised amplitudes appearing on the right-hand side of \refeq{eq:factLloop} can be obtained from the $n$-loop form factors for the production and decay of the vector boson $V$.
The factorisable corrections in \refeq{eq:factLloop} correspond to approximate the first term in \refeq{eq:PoleExp} with $\mathcal{R}(m_V^2)/(\OFFp_V^2-\mu_V^2)$.

Starting from the one-loop order, additional non-factorisable contributions, 
\begin{align}
    \M_{\nf}\equiv \left\{ \M - \M_{\fact} \right\}_{p_V^2\to  m_V^2} \,,
\end{align}
arise from interferences between the subprocesses in \refeq{eq:subprocess}.
They are formally defined as the difference between the full amplitude and the factorisable contribution~\eqref{eq:factLloop}, evaluated by taking the on-shell limit $(p_V^2\to m_V^2)$, after performing the loop integration, whenever it does not lead to a singularity.
Note that, in this limit, power-suppressed terms in the decay width $\Gamma_V$ are neglected.

It should be noted that the distinction between factorisable and non-factorisable corrections holds at the amplitude level and cannot always be attributed to individual Feynman diagrams, as we will discuss in the remaining part of this section.
Therefore, a proper analysis based on the method of regions \cite{Beneke:1997zp,Smirnov:2002pj} should be employed. 
This analysis relies on the systematic expansion of the off-shell $n$-loop virtual amplitude in terms of a small parameter \( \lambda \sim \Gamma_V/ m_V \), whose choice is justified by the fact that $\Gamma_V\ll m_V$. 
In the power counting, only the leading terms in $\lambda$ are retained. 
The loop integrals that arise in our case receive non-vanishing contributions from \textit{hard} ($k \sim \mathcal{O}(\lambda^0)$) and \textit{soft} ($k \sim \mathcal{O}(\lambda)$) regions.
\footnote{In principle, a collinear momentum scaling must also be taken into account. However, we find—consistent with Ref.\cite{Dittmaier:2014qza}—that contributions from collinear regions lead to scaleless integrals or cancel pairwise between diagrams, and thus do not affect the final result. We therefore do not consider these regions further in the remainder of this paper.}
\begin{figure}[t]
    \includegraphics[width=\textwidth]{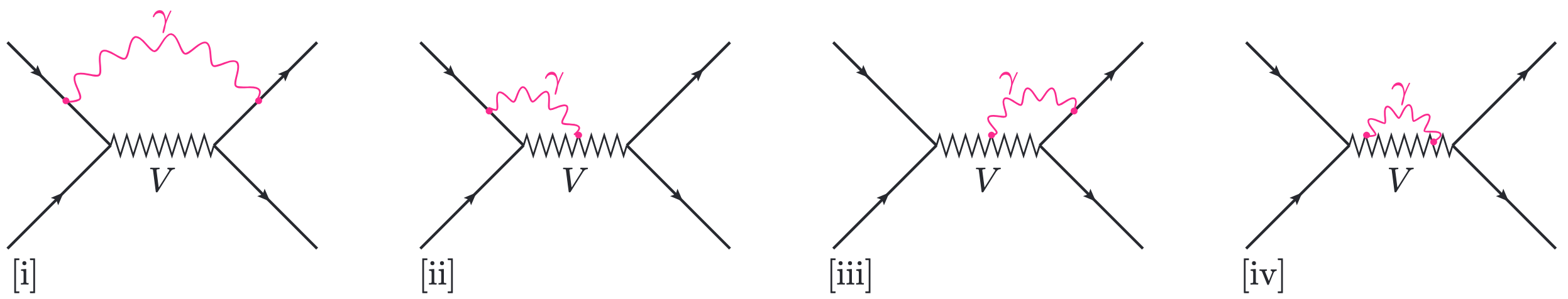}
    \caption{Representative diagrams for the four classes of one-loop non-factorisable corrections: [i] initial-final, [ii] initial-resonance, [iii] resonance-final, [iv] resonance-resonance.}
    \label{fig:nf1L}
\end{figure}

At the one-loop order, it is quite straightforward to verify that the exchange of a massive EW boson or a highly energetic photon contributes exclusively to the factorisable part, at leading power in $\lambda$.
On the contrary, the non-factorisable corrections arise from the exchange of a soft photon between the initial and final states, or between the external legs and the resonance itself (if charged).
Four classes of diagrams can therefore be identified (see Figure \ref{fig:nf1L}): initial-final ($if$), initial-resonance ($ir$), resonance-final ($rf$), and resonance-resonance ($rr$) contributions.
Among these, only the first type is relevant if the resonance $V$ is electrically neutral.
While diagrams connecting the initial- and final-state legs ($if$) give a leading non-vanishing contribution only in the soft region, diagrams in which the resonance propagator appears simultaneously in a loop and as a tree-level line ($ir$, $rf$, $rr$) contribute to both factorisable and non-factorisable corrections.
\begin{figure}[t]
    \includegraphics[width=\textwidth]{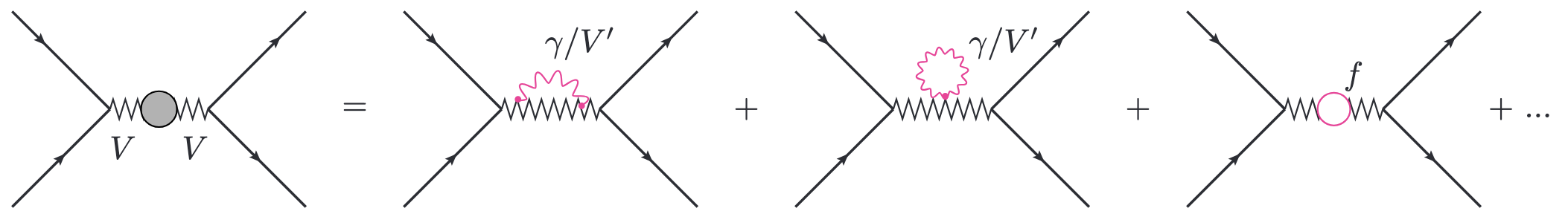}
    \caption{Representative diagrams for the one-loop corrections to the resonance propagator.}
    \label{fig:propcorr}
\end{figure}
\noindent
We also comment on the one-loop corrections to the resonance propagator shown in Figure~\ref{fig:propcorr}. 
In the soft region, only the first diagram with a photon exchange gives a leading non-factorisable contribution, corresponding to [iv] in Figure \ref{fig:nf1L}.
The remaining diagrams are either power-suppressed or vanish in dimensional regularisation. 
In the hard region, all diagrams in Figure~\ref{fig:propcorr} produce a non-vanishing contribution which is more leading than the expected $\mathcal{O}\left(\frac{1}{\lambda}\right)$ behaviour of the tree-level amplitude.
However, these {\it superleading} terms do not affect the UV renormalised amplitude, since in the OS scheme they are exactly cancelled by the vector-boson wave-function and mass counterterms (see the discussion in Ref.~\cite{Falgari:2010sf}).

\begin{figure}[h]
    \centering
      \includegraphics[width=\textwidth]{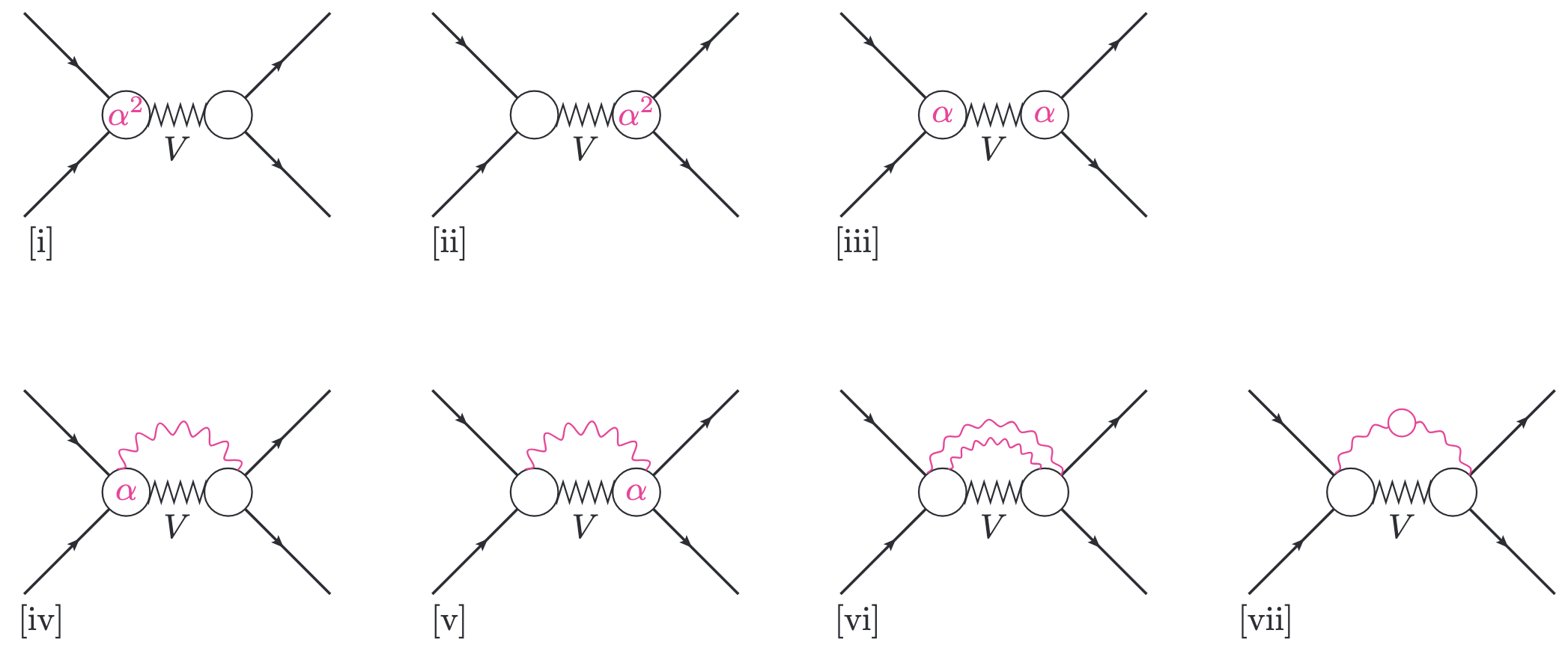}
    \caption{Classification of corrections at two-loop order: [i] two-loop correction to the on-shell production, [ii] two-loop correction to the on-shell decay, [iii] the one-loop correction to both the on-shell production and the decay, [iv] and [v] products of one-loop factorisable and one-loop non-factorisable corrections, [vi] and [vii] genuine two-loop non-factorisable corrections.}    
    \label{fig:nf2L}
\end{figure}

A similar analysis can be carried out at two-loop order. 
According to the factorisation properties of each sub-amplitude at leading power in the on-shell limit,
we can identify three categories of corrections contributing to the two-loop amplitude in PA:
\begin{enumerate}
    \item[(a)] \textbf{two-loop factorisable corrections} $(\M_{\fact}^{(2)})$, which can be constructed as in \refeq{eq:factLloop} for $n=2$, starting from the on-shell polarised amplitudes associated with the production and decay subprocesses in \refeq{eq:subprocess}. 
    They include two-loop \(\mathcal{O}(\alpha^2)\) corrections to either production or decay form factors (see diagrams [i] and [ii] in Figure \ref{fig:nf2L}) and one-loop \(\mathcal{O}(\alpha)\) corrections to both production and decay subprocesses separately (see diagram [iii] in Figure \ref{fig:nf2L});
    \item[(b)] \textbf{one-loop factorisable times one-loop non-factorisable corrections} $(\M_{\fact\otimes \nfbis}^{(2)})$. They include one-loop \(\mathcal{O}(\alpha)\) corrections to either on-shell production or decay form factors dressed with
    a soft-photon exchange between production and decay subprocesses (see diagrams [iv] and [v] in Figure \ref{fig:nf2L}).
    \item[(c)] \textbf{genuine two-loop non-factorisable corrections} $(\M_{\nfbis}^{(2)})$, which arise either from two soft-photon exchanges between production and decay subprocesses (see diagram [vi] in Figure \ref{fig:nf2L}) or from a one-loop fermionic and bosonic correction to a single soft-photon exchange (see diagram [vii] in Figure \ref{fig:nf2L}).
\end{enumerate}
In the computation of the two-loop non-factorisable corrections ((b) and (c)) detailed in Sec.~\ref{sec:twoloop}, we typically organise the contributing diagrams into classes according to the number and type of external legs and/or resonances to which the soft photon(s) is attached, rather than treating individual Feynman diagrams separately.

When both photons exchanged between the production and decay subprocesses are soft, the corresponding amplitude factorises into a universal correction factor multiplying the Born-level amplitude in PA. 
After combining all diagrams within the same class, the resulting correction factor can be expressed as a two-loop soft integral that contains only a single resonance propagator with coupled dependence on the two loop momenta. 
These eikonal-type integrals can be evaluated by direct analytic integration performed iteratively, loop by loop.
A detailed discussion will be presented in Sec.~\ref{sec:twoloop}.

Additional contributions arise when a single photon, exchanged between different subprocesses, is soft (see Secs. \ref{sec:fermionic} and \ref{sec:bosonic}).
Special care is required when the (second) hard loop corresponds to a self-energy correction to the resonance $V$ (see Figure~\ref{fig:propcorr}) dressed by the soft photon itself. 
At first glance, these contributions do not fall into any of the categories (a), (b), or (c) at bare level. 
Nevertheless, they do not contribute to the UV renormalised amplitude. 
As in the one-loop case, they cancel against corresponding diagrams with insertions of the OS mass and wave-function counterterms associated with the resonance.
The only exception arises when the soft photon couples to a fermionic or bosonic self-energy correction of the neutral resonance. 
In this case, the cancellation already occurs at the bare level after summing over all possible attachments of the soft photon to the external particles. This behaviour is a direct consequence of charge conservation, together with the fact that the corresponding soft integral depends kinematically only on the resonance mass.
\\
\\
The workflow adopted for our computation is as follows. 
We generate all relevant Feynman diagrams using automated tools such as \texttt{FeynArts} \cite{Hahn:2000kx} and \texttt{QGRAF} \cite{Nogueira:1991ex}. 
The corresponding amplitudes are reconstructed and simplified using \texttt{FeynCalc} \cite{Shtabovenko:2024aum}. 
We then apply the method of regions to systematically extract the non-factorisable contributions and express them in terms of soft integrals at both one- and two-loop order. 
Throughout the calculation, we work in the unitary gauge. 
However, the relevant parts of the computation have been repeated in the Feynman gauge to prove the gauge independence of the final results.
The UV renormalisation is performed in the OS scheme for the mass and wave-functions, and in the $\MSbar$ scheme for the electric charge (see Sec.~\ref{sec:results_UVren}).

We conclude this section with a few remarks on a different approach to the one adopted here. 
As discussed above, our treatment of radiative corrections in the presence of unstable particles relies on the method of regions and on an analysis of the relevant Feynman diagrams. 
Given the hierarchical structure of scales in the problem, $\Gamma_V \ll m_V$, an alternative approach is provided by Effective Field Theory (EFT) methods~\cite{Beneke:2003xh,Beneke:2004km,Beneke:2015vfa}.
In this framework, the analysis similarly begins by separating the relevant momentum regions in the Feynman integrals. 
This separation is then used to construct an effective Lagrangian in which the corresponding degrees of freedom are explicitly disentangled at the field level. The hard modes, which are responsible for the factorisable corrections in our approach, are integrated out and encoded in the proper matching coefficients, while the soft modes represent the propagating degrees of freedom and lead to the non-factorisable corrections.
The EFT formulation provides a transparent field-theoretical interpretation of the pole approximation and, in addition, enables the resummation of specific classes of higher-order contributions. This has been successfully exploited, for instance, in the description of $e^+e^- \to 4$-fermion production near the $WW$ threshold \cite{Actis:2008rb}.
At the same time, the EFT approach has intrinsic limitations. Indeed, given that some of the degrees of freedom are integrated out, a (fully) differential description of the final state may not always be possible.

\subsection{The one-loop calculation} \label{sec:oneloop}   
 
As already discussed in the previous section, the non-factorisable corrections for the process in \refeq{eq:process} arise from soft wide-angle photon exchanges between production and decay stages.
As a result, the one-loop coefficient of the non-factorisable corrections can be recast in the following form
\begin{align}
    2 \mathrm{Re}\left\{\M_{\PA}^{(0)*}\M^{(1)}_{\nf}\right\} = 2 \mathrm{Re}\!\left(\delta_{\nfbis}^{(1)}\right)\left|\M_{\PA}^{(0)}\right|^2\,,
\end{align}
where $\delta_{\nfbis}^{(1)}$ is a universal function depending on the kinematics of the external particles.
At one-loop order, $\delta_{\nfbis}^{(1)}$ receives contributions from four classes of corrections depicted in Fig. \ref{fig:nf1L}. Their expression is given by 
\begin{align}
    \delta_{if}^{(1)} = &i (4\pi)^2(\sigma_i e_i)(\sigma_fe_f ) (4\ONp_i\cdot \ONp_f) \;K_V\;  \D[\{V,i,f\};1,1]
    \label{eq:1lif}
    \\
    \delta_{rf}^{(1)} = &-i(4\pi )^2(\sigma_f e_f) (\sigma_Ve_V) (4\ONp_V\cdot \ONp_f) \;\C[\{V,f\};1,1]
    \label{eq:1lrf}
    \\
    \delta_{ir}^{(1)} = &i(4\pi)^2(\sigma_i e_i) (\sigma_Ve_V) (4\ONp_V\cdot \ONp_i) \;\C[\{V,i\};1,1]
    \label{eq:1lir}
    \\
    \delta_{rr}^{(1)} = & -i (4\pi)^2 (\sigma_Ve_V)^2 \;\frac{4m_V^2}{K_V}\;\B[\{V\};1,1]\,,
    \label{eq:1lrr}
\end{align}
where $\D,\C,$ and $\B$ are soft scalar integrals whose definitions and explicit results are reported in Appendix \ref{app:1}.
Here, the set $\{\ONp_n\}$ denotes the external momenta after on-shell projection with $\ONp_V^2=m_V^2$, $K_V = \OFFp_V^2 -\mu_V^2$ is the off-shell propagator of the resonance $V$, and to each particle we assign a relative electric charge $e_n$ satisfying global
\begin{equation}
    \sum_{n\in\mathcal{I} \cup \mathcal{F}}\sigma_n e_n = 0\,
\end{equation}
and local 
\begingroup
\allowdisplaybreaks
\begin{align}
     \sum_{i \in\mathcal{I} }\sigma_i e_i  + \sigma_Ve_V= 0\,,\qquad\qquad\sum_{f \in\mathcal{F} }\sigma_f e_f  - \sigma_Ve_V= 0\,
\end{align}
\endgroup
charge conservation. In the above equations, $\mathcal{I} =\{f_1,\bar{f}_2\}$ and  $\mathcal{F} =\{f_3,\bar{f}_4\}$ denote the set of initial- and final-state particles, respectively.
We define $\sigma_n=+1$ if $f_n$ is an incoming particle or an outgoing anti-particle, and $\sigma_n=-1$ otherwise. 
In our convention, the resonance $V$ is always treated as an outgoing state, with $\sigma_{W^\pm}=\pm 1$ and $e_{W^\pm}=-1$.
Once we sum over all possible combinations of external legs contributing to \refeqs{eq:1lif}-(\ref{eq:1lrr}), we obtain 
a gauge-invariant result whose expression to all orders in $\epsilon$ is 
\begingroup
\allowdisplaybreaks
\begin{align}
    \delta_{\nfbis}^{(1)}& =  S_\epsilon
    \left(\frac{\mu_0^2}{m_V^2}\right)^{\epsilon}\left(\frac{-K_V-i0_+}{m_{V}^2}\right)^{-2\epsilon}
    e^{\gamma_E\epsilon}\Gamma[1+2\epsilon]\Gamma[1-\epsilon]
    \Bigg[
    \frac{2}{\epsilon(1-2\epsilon)}(\sigma_Ve_V)^2 \notag \\
    &-\frac{\sigma_Ve_V}{\epsilon^2}\sum_{i\in\mathcal{I}}\sigma_ie_i
    +\frac{\sigma_Ve_V}{\epsilon^2}\sum_{f\in\mathcal{F}}\sigma_fe_f
    + \frac{2}{\epsilon^2}\sum_{\substack{i\in\mathcal{I}\\f\in\mathcal{F}}}\sigma_ie_i \sigma_fe_f\,
   {}_2F_1\left(1,-\epsilon;1-\epsilon;1-\frac{m_V^2}{(2\ONp_i\cdot \ONp_f)}\right)
    \Bigg]\,,
    \label{eq:deltanf1L}
\end{align}
\endgroup
where $\mu_0$ is the dimensional-regularization scale and $S_\epsilon = (4\pi)^\epsilon e^{-\gamma_E\epsilon}$.
By expanding the result in \refeq{eq:deltanf1L} as a Laurent series in $\epsilon$ and using the charge-conservation relations, we obtain
\begin{align}
    \delta_{\nfbis}^{(1)} &=S_\epsilon
    \left(\frac{\mu_0^2}{m_V^2}\right)^{\epsilon}\left(\frac{-K_V-i0_+}{m_{V}^2}\right)^{-2\epsilon}
    \!\!e^{2 \gamma_E\epsilon}\,\Gamma[1+2\epsilon] \notag \\
    &\hspace{2cm}\times\sum_{\substack{i\in\mathcal{I}\\f\in\mathcal{F}}}\sigma_ie_i\sigma_fe_f \left[
\frac{2}{\epsilon}\left(\log\left(\frac{m_V^2}{2\ONp_i\cdot \ONp_f}\right)-1\right)-2\text{Li}_2\left(1-\frac{m_V^2}{2\ONp_i\cdot \ONp_f}\right)-4+\mathcal{O}(\epsilon)
\right]\,,
\label{eq:deltanf1Lexp}
\end{align}
which is consistent with the results in the literature \cite{Dittmaier:2001ay,Denner:2019vbn}.
Note also that the kinematically independent terms cancel identically in the case of a neutral resonance $(V = Z)$ due to charge conservation.
We conclude the discussion of the one-loop non-factorisable corrections by highlighting the following important features:
\begin{itemize}
    \item Even though individual contributions in \refeqs{eq:1lif}-(\ref{eq:1lrr}) contain double poles in \(\epsilon\), the sum of all relevant contributions contains at most single poles, confirming the purely soft nature of the non-factorisable corrections.
    \item $\delta_{\nfbis}^{(1)}$ shows a logarithmic dependence on the off-shell resonance propagator, which eventually implies the presence of logarithmic terms in the width $\Gamma_V$ upon integration over the virtuality of the resonance. 
    Only a single power of $\log(\Gamma_V)$ can arise per loop order, as observed in Ref.~\cite{Falgari:2013gwa}, and its coefficient is fully determined by the coefficient of the single $\epsilon$ pole.
\end{itemize}

\subsection{The two-loop calculation} \label{sec:twoloop}  
In this section, we present the computation of the genuine two-loop non-factorisable corrections, which involve two distinct classes of diagrams. We first consider, in Sec.~\ref{sec:two_soft-photons}, contributions arising from diagrams with two soft photons exchanged between the production and decay subprocesses. We then turn to the second class of diagrams involving a single soft-photon exchange. The fermionic contributions are analysed first, in Sec.~\ref{sec:fermionic}, while the non-abelian corrections are discussed in the final subsection, Sec.~\ref{sec:bosonic}.

\subsubsection{Two soft-photon exchanges} \label{sec:two_soft-photons} 
The categorisation of the double soft integrals contributing to the two-loop non-factorisable corrections is more involved and can be illustrated through a few representative examples.
\begin{figure}[t]
    \centering
    \includegraphics[width=0.5\textwidth]{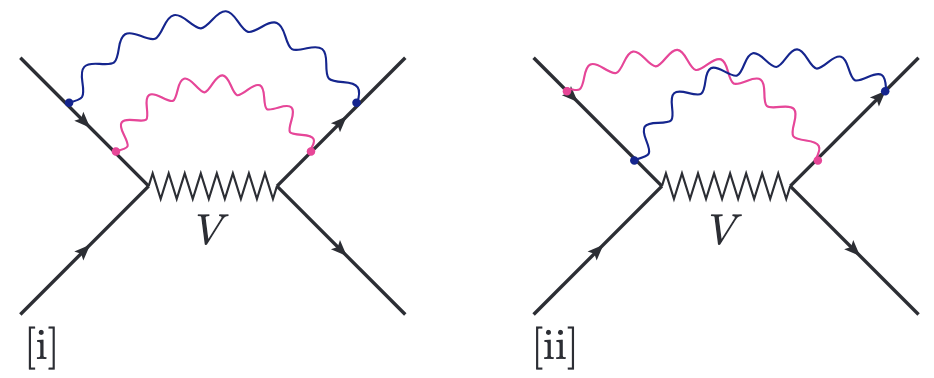}
    \caption{Example of two-loop diagrams involving a pair of virtual soft photons exchanged between the same initial- and final-state fermions.}
    \label{fig:2Lifif}
\end{figure}

We start by considering diagrams in which the two virtual photons are exchanged between the same pair of initial- and final-state legs (see Figure \ref{fig:2Lifif}).
This represents the most complicated integral family we will encounter, as it contains the highest number of propagators.
From the power-counting analysis, the only non-vanishing contribution to the non-factorisable corrections arises when both loop momenta $(l_1,l_2)$ are soft.
All other momentum regions are suppressed in $\lambda$.
In this double-soft configuration, each diagram is associated with a two-loop eikonal integral, where a few linear propagators depend on both loop momenta in a coupled way.
To decouple the loop integrations and recast them as two nested one-loop integrals, we need to sum over all diagrams contributing to the category under consideration. 
Moreover, a crucial trick is the symmetrisation of the integrand function with respect to $l_1$ and $l_2$, which leads to 
\begingroup
\allowdisplaybreaks
\begin{align}
    \delta_{if\otimes if}^{(2)} 
    &=\frac{1}{2} \int\frac{ d^dl_1 d^dl_2}{(2\pi)^{2d}}\left[\delta^{[\text{i}]}(l_1,l_2)+\delta^{[\text{ii}]}(l_1,l_2)
    +\delta^{[\text{i}]}(l_2,l_1)+\delta^{[\text{ii}]}(l_2,l_1)\right]
    \notag\\
    &=
    - \frac{1}{2} \!
   \int  \! \frac{d^dl_1d^dl_2}{(2\pi)^{2d}} \frac{(4\pi\mu_0^{\epsilon})^4
   (e_ie_f)^2(4\OFFp_i\cdot \OFFp_f)^2\;K_V}{l_1^2l_2^2(2l_1\cdot \OFFp_i+i0_+)(2l_1\cdot \OFFp_f+i0_+\!)(2l_2\cdot \OFFp_{i}+i0_+\!)(2l_2\cdot \OFFp_{f}+i0_+) (2l_1\cdot \OFFp_{V}+2l_2\cdot \OFFp_{V}+K_V)}\,,
\end{align}
\endgroup
where $\delta^{[\text{i}]}$ and $\delta^{[\text{ii}]}$ refer to the soft integrands associated with the first and second diagram in Figure \ref{fig:2Lifif}, respectively, after stripping $\M_{\PA}^{(0)}$ in \refeq{eq:M0PA}.
The only coupled dependence on $l_1$ and $l_2$ remains in the resonance propagator.
We first integrate over $l_2$ by treating $l_1$ as an external momentum. 
Using the box integral with linear propagators in \refeq{eq:D} of Appendix \ref{app:1}, with $a_0=a_1=1$, we obtain
\begingroup
\allowdisplaybreaks
\begin{align}
    \delta_{if\otimes if}^{(2)} 
    &=
    iS_\epsilon (4\pi\mu_0^{2\epsilon})^2
   (e_ie_f)^2
   \frac{e^{\gamma_E \epsilon}\Gamma[1-\epsilon]\Gamma[1+2\epsilon]}{\epsilon^2} e^{-2i \pi \epsilon} (4\OFFp_i\cdot \OFFp_f)
   (\OFFp_V^2)^\epsilon\;K_V\;    \notag\\
   &\times {}_2F_1\!\left(1,-\epsilon; 1-\epsilon; 1-\frac{(2\OFFp_i\cdot \OFFp_V)(2\OFFp_f\cdot \OFFp_V)}{\OFFp_V^2 (2\OFFp_i\cdot \OFFp_f)}\right)\!
   \int \!\frac{d^dl_1}{(2\pi)^{d}}\frac{1}{l_1^2(2l_1\cdot \OFFp_i+i0_+)(2l_1\cdot \OFFp_f+i0_+)(2l_1\cdot \OFFp_V+K_V)^{1+2\epsilon}} \,,
\label{eq:deltaifif2}
\end{align}
\endgroup
where the remaining integral over $l_1$ is identical to the one-loop initial-final box integral $\D[\{V,i,f\};1,1+2\epsilon]$ up to the dimension shift in the resonance propagator. 
After performing the integral over $l_1$, we obtain
\begingroup
\allowdisplaybreaks
\begin{align}
    \delta_{if\otimes if}^{(2)} 
    &=
    S_\epsilon^2 \hspace{-2pt}
    \left(\!\frac{\mu_0^2}{m_V^2} \!\right)^{\!\!2\epsilon}\!\! \left(\frac{-K_V-i0_+}{m_V^2}\right)^{\!\!-4\epsilon}
   \frac{2e^{2\gamma_E\epsilon}\Gamma[1+4\epsilon]}{\epsilon^4}
   \!\!
   \left[(e_ie_f) \Gamma[1-\epsilon]
   {}_2F_1\!\!\left(1,-\epsilon; 1-\epsilon ;1-\frac{m_V^2}{(2\ONp_i\cdot \ONp_f)}\right)\right]^2\,,
   \label{eq:ififexp}
\end{align}
\endgroup
where the off-shell momenta $\{\OFFp_n\}$ have been replaced with their on-shell projected counterparts $\{\ONp_n\}$, and we used the fact that $2\ONp_i\cdot \ONp_V = 2\ONp_f\cdot \ONp_V = m_V^2$ for the massless scattering we are considering.
The result is valid at all orders in the dimensional regularisation parameter $\epsilon$.
We highlight that this contribution can be written in a fully factorised form 
\begingroup
\allowdisplaybreaks
\begin{align}
    \delta_{if\otimes if}^{(2)} & =\frac{\Gamma[1+4\epsilon]}{\Gamma[1+2\epsilon]^2} \,\frac{1}{2}\!\left(\delta_{if}^{(1)}\right)^2\,
    \label{eq:nf2Lifif}
\end{align}
\endgroup
as the square of the one-loop initial-final correction $\delta_{if}^{(1)}$ weighted by the ratio of $\Gamma[1+2n\epsilon]/\Gamma[1+2\epsilon]^n$, where $n$ stands for the loop order. 
The factor $1/2$ arises from the symmetrisation over the loop momenta.

\begin{figure}[t]
    \centering
    \includegraphics[width=0.8\textwidth]{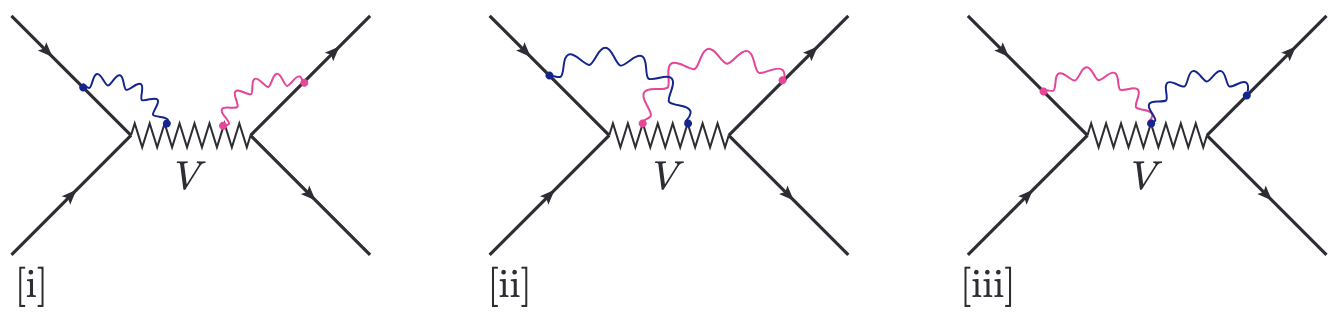}
    \caption{Example of two-loop $ir\otimes rf$ diagrams where one soft photon is exchanged between resonance- and initial-state, and the other between resonance- and final-state particles.}
    \label{fig:2Lirrf}
\end{figure}

We now consider diagrams where at least one virtual photon is attached to the resonance, contributing to more than one class of two-loop corrections identified in Sec.\ref{sec:preliminaries}.
In particular, here we will focus on a subset of those diagrams, labelled as $ir\otimes rf$, where one photon is exchanged between resonance- and initial-state, and the other between resonance- and final-state particles (see Figure~\ref{fig:2Lirrf}).
Before presenting the result of the genuine two-loop non-factorisable corrections, $\delta^{(2)}_{ir\otimes rf}$, we briefly discuss the corresponding power-counting analysis.

The third diagram in Figure~\ref{fig:2Lirrf}, which contains the four-point gauge interaction vertex, is subleading in all photon momentum regions and can therefore be neglected.
The second diagram yields a leading contribution only in the double-soft region, i.e.\ where both loop momenta have a soft scaling $l\sim m_V \lambda$. 
In contrast, the first diagram receives contributions from several momentum regions. More precisely, if both loop momenta are hard, the corresponding amplitude factorises into the product of one-loop corrections to the on-shell production and decay subprocesses (see \refeq{eq:subprocess}). 
If the photon exchanged between the initial state and the resonance is hard while the other photon is soft, the amplitude factorises into the one-loop non-factorisable correction $\delta_{rf}^{(1)}$ multiplied by the one-loop correction to the on-shell production subprocess. 
Similar factorisation  $(\delta_{ir}^{(1)}\M_{\fact}^{(1)})$ holds if the scalings of the two photon momenta are interchanged.
In the double-soft region, the amplitude factorises into the Born matrix element in PA~\eqref{eq:M0PA} times a two-loop eikonal integral.
Combining the latter with an analogous contribution from the second diagram in Figure~\ref{fig:2Lirrf}, we obtain the two-loop coefficient
\begingroup
\allowdisplaybreaks
\begin{align}
     \delta_{ir\otimes rf}^{(2)} &= -(4\pi\mu_0^{\epsilon})^4 (\sigma_i e_i) (\sigma_f e_f) (\sigma_Ve_V)^2 (4\OFFp_V\cdot \OFFp_i)(4\OFFp_V\cdot \OFFp_f)
       \notag\\
   &\times \int \!\frac{d^dl_1d^dl_2}{(2\pi)^{2d}}\frac{(2l_1\cdot \OFFp_V+K_V)+(2l_2\cdot \OFFp_V+K_V)}{l_1^2 l_2^2(2l_1\cdot \OFFp_i+i0_+)(2l_1\cdot \OFFp_V+K_V)(2l_2\cdot \OFFp_f+i0_+)(2l_2\cdot \OFFp_V+K_V)(2l_1\cdot \OFFp_V+2l_2\cdot \OFFp_V + K_V)}\, ,
\end{align}
\endgroup
where the double integration can be carried out following a similar procedure described for $\delta_{if\otimes if}^{(2)}$.
Details on the analytic results of the necessary soft integrals are provided in Appendix~\ref{app:1}. 
Finally, a factorisation
\begin{align}
    \delta_{ir\otimes rf}^{(2)}=& \frac{\Gamma[1+4\epsilon]}{\Gamma[1+2\epsilon]^2}\delta_{ir}^{(1)}\delta_{rf}^{(1)}\,
\end{align}
equivalent to \refeq{eq:nf2Lifif} is obtained in terms of one-loop results $\delta_{rf}^{(1)}$ and $\delta_{ir}^{(1)}$ given in \refeqs{eq:1lrf} and (\ref{eq:1lir}), respectively.
The total contribution to the two-loop amplitude in PA from the class of diagrams in Figure~\ref{fig:2Lirrf} is found to be
\begin{align}
    \M_{ir\otimes rf}^{(2)} = \frac{1}{K_V}\left\{\M_{\produc}^{(1)}\M_{\dec}^{(1)}
    + \delta_{rf}^{(1)}\M_{\produc}^{(1)}\M_{\dec}^{(0)} +\delta_{ir}^{(1)} \M_{\produc}^{(0)}\M_{\dec}^{(1)}+\frac{\Gamma[1+4\epsilon]}{\Gamma[1+2\epsilon]^2}\delta_{ir}^{(1)}\delta_{rf}^{(1)} \M_{\produc}^{(0)}\M_{\dec}^{(0)}\right\}\,,
\end{align}
where the first term in the curly bracket contributes to the two-loop factorisable corrections.

A careful treatment is required for the class of diagrams ${rr\otimes ab}$ obtained by dressing the one-loop diagrams in Figure~\ref{fig:propcorr} with an additional photon exchanged between legs $a$ and $b$, thereby connecting the production and decay subprocesses.
If both loop momenta are soft, a leading contribution to the genuine two-loop non-factorisable corrections arises. This contribution can be recast in the same factorised form discussed in the previous examples.
When the loop momentum associated with the $rr$ subdiagram is soft while the second loop momentum is hard, the corresponding contribution is either power-suppressed or reduces to the product $\delta_{rr}^{(1)} \mathcal{M}_{\text{fact}}^{(1)}$.
Finally, if the loop momentum in the $rr$ subdiagram is hard---independently of the scaling of the second loop momentum---the diagrams do not contribute to the UV renormalised amplitude, as already discussed in Sec.~\ref{sec:preliminaries}.

Extending the previous examples to all two-loop diagrams involving a pair of soft photons exchanged between the production and decay subprocesses, we denote the corresponding corrections by $\delta^{(2)}_{ab\otimes cd}$. Here, the subscript $ab$ ($cd$) identifies the pair of edges to which the first (second) soft-photon loop is attached.
The complete set of results is given by
\begingroup
\allowdisplaybreaks
\begin{align}
   &\delta_{if\otimes i'f'}^{(2)} =-(4\pi)^4\!\!\left(\!1-\frac{1}{2}\delta_i^{i'}\delta_f^{f'}\!\right)\!
   (\sigma_ie_i)(\sigma_{i'}e_{i'})(\sigma_fe_f)(\sigma_{f'}e_{f'})(4\ONp_i\cdot \ONp_f)(4\ONp_{i'}\cdot \ONp_{f'})\;K_V \;
   \DD[\{i,f\},\{i',f'\},V] \label{eq:ifixfx}
   \\
    &\delta_{if\otimes i'r}^{(2)} = -(4\pi)^4
   (\sigma_ie_i)(\sigma_{i'}e_{i'})(\sigma_fe_f) (\sigma_Ve_V) (4\ONp_i\cdot \ONp_f)(4\ONp_{i'}\cdot \ONp_{V})\;K_V \;
   \DC[\{i,f,V\},\{i'\},V] 
   \\
   &\delta_{if\otimes rf'}^{(2)} = (4\pi)^4
   (\sigma_ie_i)(\sigma_fe_f)(\sigma_{f'}e_{f'}) (\sigma_Ve_V) (4\ONp_i\cdot \ONp_f)(4\ONp_{f'}\cdot \ONp_{V})\;K_V \;
   \DC[\{i,f,V\},\{f'\},V] 
   \\
   &\delta_{if\otimes rr}^{(2)} =(4\pi)^4 (\sigma_i e_i)(\sigma_f e_f) (\sigma_Ve_V)^2 (4\ONp_i\cdot \ONp_f) (4m_V^2)\;K_V\;
   \DB[\{i,f,V\},\{\},V;2]
   \\
   &\delta_{ir\otimes i'r}^{(2)} = -(4\pi)^4\left(1-\frac{1}{2}\delta_i^{i'}\right)(\sigma_i e_i) (\sigma_{i'} e_{i'}) (\sigma_Ve_V)^2 (4\ONp_V\cdot \ONp_i)(4\ONp_V\cdot \ONp_{i'})
    \left(\CC[\{i,V\},\{i'\},V]+\CC[\{i',V\},\{i\},V]\right)
    \\
    &\delta_{ir\otimes rf}^{(2)}= (4\pi)^4 (\sigma_i e_i) (\sigma_f e_f) (\sigma_Ve_V)^2 (4\ONp_V\cdot \ONp_i)(4\ONp_V\cdot \ONp_f)
   \left(\CC[\{i,V\},\{f\},V]+\CC[\{f,V\},\{i\},V]\right)
   \\
    &\delta_{ir\otimes rr}^{(2)} = (4\pi)^4(\sigma_i e_i)(\sigma_Ve_V)^3(4\ONp_i\cdot \ONp_V)\frac{4m_V^2}{K_V}
    \left( K_V \CB[\{i,V\},\{\},V;2] 
    + \CB[\{i\},\{V\},V]
    + \CB[\{i,V\},\{\},V;1] \right)
    \\
    &\delta_{rf\otimes rf'}^{(2)} = -(4\pi)^4\!\!\left(\!1-\frac{1}{2}\delta_f^{f'}\!\right)\!(\sigma_fe_f) (\sigma_{f'} e_{f'}) (\sigma_Ve_V)^2 (4\ONp_V\cdot \ONp_f)(4\ONp_V\cdot \ONp_{f'}\!)\!
    \left(\CC[\{f,V\},\{f'\},V]+\CC[\{f',V\},\{f\},V]\right)
    \\
    &\delta_{rf\otimes rr}^{(2)} = -(4\pi)^4(\sigma_f e_f) (\sigma_Ve_V)^3 (4\ONp_V\cdot \ONp_f)\frac{4m_V^2}{K_V}
    \left( K_V \CB[\{f,V\},\{\},V;2] 
    +  \CB[\{f\},\{V\},V]
    +  \CB[\{f,V\},\{\},V;1] \right)
    \\
   &\delta_{rr\otimes rr}^{(2)} = -(4\pi)^4 (\sigma_Ve_V)^4 \left(\frac{4m_V^2 }{K_V}\right)^2
    \left(K_V\BB[\{V\},\{\},V; 2] + 2\BB[\{V\},\{\},V; 1]\right)\label{eq:rrrr}\,.
\end{align}
\endgroup
In the previous expressions, the indices are defined as $i,i'\in \mathcal{I}$ and $f,f'\in \mathcal{F}$, while $r$ refers to the resonance vector $V$. 
Explicit results and conventions for the two-loop soft integrals $\DD,\DC,\DB,\CC,\CB,\BB$ are reported in Appendix~\ref{app:1}.
We remind the reader that only the first contribution in \refeq{eq:ifixfx} is present if $V$ is a neutral vector boson.

\subsubsection{Single soft-photon exchange: Fermionic contributions}\label{sec:fermionic}
\begin{figure}[t]
    \includegraphics[width=\textwidth]{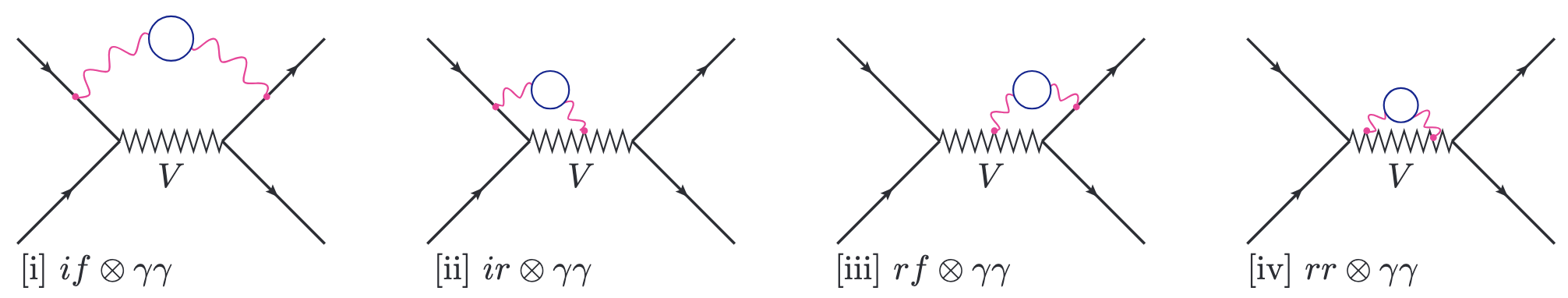}
    \caption{Non-factorisable corrections arising from a self-energy insertion on the soft-photon propagator.}
    \label{fig:2LnfAA}
\end{figure}

In this subsection, we discuss diagrams featuring a fermion-loop insertion in the soft-photon propagator (see Figure~\ref{fig:2LnfAA}). 
All other configurations (e.g.\ those contributing to $Z\gamma$ mixing) are subleading.
We denote these contributions by $\delta_{ab\otimes \gamma\gamma}^{(2)}$, where $ab$ labels the two edges to which the photon is attached, while the shorthand $\gamma\gamma$ refers to the loop correction to the photon propagator.

The leading contribution arises when the momentum $l$ is soft, independently of the scaling of the momentum $k$ in the inner loop. 
A naive power counting at the fully unintegrated level yields power-like singularities of order $1/\lambda^3$, suggesting an enhancement of the two-loop amplitude by two additional powers of $\lambda$ relative to the tree-level scaling. 
However, these singularities are spurious and reduce to the expected $1/\lambda$ behaviour after integration over $k$.
As a result, the amplitudes corresponding to these diagrams can be obtained from their one-loop counterparts by replacing the tree-level photon propagator with its one-loop correction (see Figure~\ref{fig:nf1L})
\begin{align}
    \frac{-ig^{\mu\nu}}{l^2}\to \frac{- i}{l^2} \left(i\:\Pi_{\gamma\gamma}^{\mu\nu,(1)}(l)\right)\frac{-i}{l^2 } = -\frac{i}{l^2} \left(-g^{\mu\nu}+ \frac{l^\mu l^\nu}{l^2}\right)\Pi_{\gamma\gamma}^{(1)}(l^2)\,,
\end{align}
where $\Pi_{\gamma\gamma}^{\mu\nu}(l) = (-g^{\mu\nu} l^2 + l^{\mu}l^\nu) \Pi_{\gamma\gamma}(l^2)$ and $\Pi_{\gamma\gamma}^{(1)}(l^2)$ stands for the scalar integral associated with the one-loop photon self-energy.
In the above equation, the photon's transversality is manifest as a consequence of Lorentz and gauge invariance.
The longitudinal contributions proportional to $l^\mu l^\nu$ vanish once the full gauge-invariant set of photon attachments is taken into account.
The same gauge-invariant result is therefore obtained by directly omitting the $l^\mu l^\nu$ terms in every diagram. 

The one-loop self-energy correction $\Pi_{\gamma\gamma}^{(1)}(l^2)$ receives a contribution from both massive ($\Pi_{\gamma\gamma}^{(1),F}$) and massless ($\Pi_{\gamma\gamma}^{(1),\ell}$) fermions.
We first consider $\Pi_{\gamma\gamma}^{(1),F}$, assuming that the mass $m_F$ of the heavy particle is much larger than the width of the resonance ($m_F \gg \Gamma_V$). 
After integrating out the inner loop, the result depends on $l^2$ and $m_F$. 
Since the incoming momentum is soft, we consider the $l^2 \to 0$ limit and retain only the leading contributions in the $\lambda$-expansion, yielding
\begingroup
\allowdisplaybreaks
\begin{align}
    \Pi_{\gamma\gamma}^{(1),F}(0) =   S_\epsilon e^{\gamma_E\epsilon} \Gamma[\epsilon]
    \left[
        \frac{4}{3} \sum_{j\in\{h\}} N_c^{j} e_j^2 \left(\frac{m_j^2}{\mu_0^2}\right)^{-\epsilon}
    \right]\,,
    \label{eq:PIheavyF}
\end{align}
\endgroup
where $\{h\}$ is the set of massive fermions in the theory and $N_c^j = 1\, (3)$ if $j$ is a lepton (quark). 
Combined with the rest of the computation, this leads to an absolute factorisation 
\begin{align}
    \delta_{ab \otimes \gamma\gamma}^{(2),F} = - \delta_{ab}^{(1)} \; \Pi_{\gamma\gamma}^{(1),F}(0)\,,
    \label{eq:abAAheavyF}
\end{align}
for each pair $ab$ of edges to which the soft photon is attached.

This simplification does not apply to the massless fermion case, since the inner loop integration generates a non-trivial dependence on the photon momentum $l$. Consequently, the soft limit cannot be implemented by simply setting $l^2=0$. In this case, we obtain~\footnote{We note that the phase factors $e^{-i\pi \epsilon}$ cancel out when combined with the explicit expression of the soft integrals with $a_0=1+\epsilon$ (see Appendix \ref{app:1}).}
\begingroup
\allowdisplaybreaks
\begin{align}
     &\delta_{if\otimes \gamma\gamma}^{(2),\ell} = i (4\pi)^2 S_\epsilon \mu_0^{2\epsilon } (\sigma_i e_i)(\sigma_f e_f) (4\ONp_i\cdot \ONp_f)K_V e^{-i \pi \epsilon}A(\epsilon)\frac{\beta_0}{\epsilon}\;
     \mathcal{D}[\{V,i,f\};1+\epsilon,1]\label{eq:2LifAA}
    \\
    &\delta_{rf\otimes  \gamma\gamma}^{(2),\ell} = - i(4\pi)^2 S_\epsilon \mu_0^{2\epsilon } (\sigma_V e_V) (\sigma_f e_f) (4\ONp_V\cdot \ONp_f) e^{-i \pi \epsilon}A(\epsilon)\frac{\beta_0}{\epsilon}\;
    \mathcal{C}[\{V,f\};1+\epsilon,1]\label{eq:2LrfAA}
    \\
   &\delta_{ir\otimes  \gamma\gamma}^{(2),\ell} = i (4\pi)^2 S_\epsilon \mu_0^{2\epsilon } (\sigma_i e_i) (\sigma_V e_V) (4\ONp_V\cdot \ONp_i) e^{-i \pi \epsilon}A(\epsilon)\frac{\beta_0}{\epsilon}\;
    \mathcal{C}[\{V,i\};1+\epsilon,1]\label{eq:2LirAA}
    \\
    &\delta_{rr\otimes  \gamma\gamma}^{(2),\ell} = -i (4\pi)^2 S_\epsilon \mu_0^{2\epsilon } (\sigma_Ve_V)^2\frac{4m_V^2}{K_V} e^{-i \pi \epsilon} A(\epsilon) \frac{\beta_0}{\epsilon}\;
    \mathcal{B}[\{V\};1+\epsilon,1]\,,\label{eq:2LrrAA}
\end{align}
\endgroup
where 
\begin{equation}
    A(\epsilon)=  \frac{6\;e^{\gamma_E\epsilon}\Gamma[1+\epsilon]\Gamma[2-\epsilon]^2}{\Gamma[4-2\epsilon]} = 1+\frac{5}{3}\epsilon+\mathcal{O}(\epsilon^2)\,.
\end{equation}
The one-loop coefficient of the QED $\beta$ function is given by
\begin{align}
    \beta_0 &= -\frac{4}{3}\sum_{j \in \{\ell\}} N_c^{j}  e_j^2  \,,
    \label{eq:beta0}
\end{align}
where the set $\{\ell\}$ denotes the possible massless quark and lepton species in the theory. 
Note that the one-loop scalar integrals in \refeqs{eq:2LifAA}–(\ref{eq:2LrrAA}) are identical to those in \refeqs{eq:1lif}–(\ref{eq:1lrr}), except that the first denominator appears with a shifted power, i.e.\ the propagator $l^2$ is raised to $(1+\epsilon)$.
This shift arises from the integration of the bubble subdiagram and follows directly from its mass dimension.
We observe that the effect of the light-fermion contribution at two-loop order $(\delta_{ab\otimes \gamma\gamma}^{(2),\ell})$ can be captured by substituting $\alpha_0$ with the $d$-dimensional effective (running) coupling
\begin{align}
    \alpha(l^2) \equiv\alpha_0 \left(1+ \frac{\alpha_0}{4\pi}S_\epsilon A(\epsilon)\frac{\beta_0}{\epsilon}\left(\frac{\mu_0^2}{-l^2}\right)^{\!\epsilon} \right)
\end{align}
in the one-loop non-factorisable integral
\begin{align}
  \frac{\alpha_0}{4\pi} \delta_{ab}^{(1)} \equiv \frac{S_\epsilon\mu_0^{2\epsilon}}{4\pi} \int \frac{d^dl}{(2\pi)^d} \;\alpha_0\; \delta_{ab}^{(1)'}(l) \,,
\end{align}
where $\delta_{ab}^{(1)'}$ symbolically denotes the eikonal integrand.

\subsubsection{Single soft-photon exchange: Non-Abelian contributions}\label{sec:bosonic} 
\begin{figure}[t]
   \centering
    \includegraphics[width=0.6\textwidth]{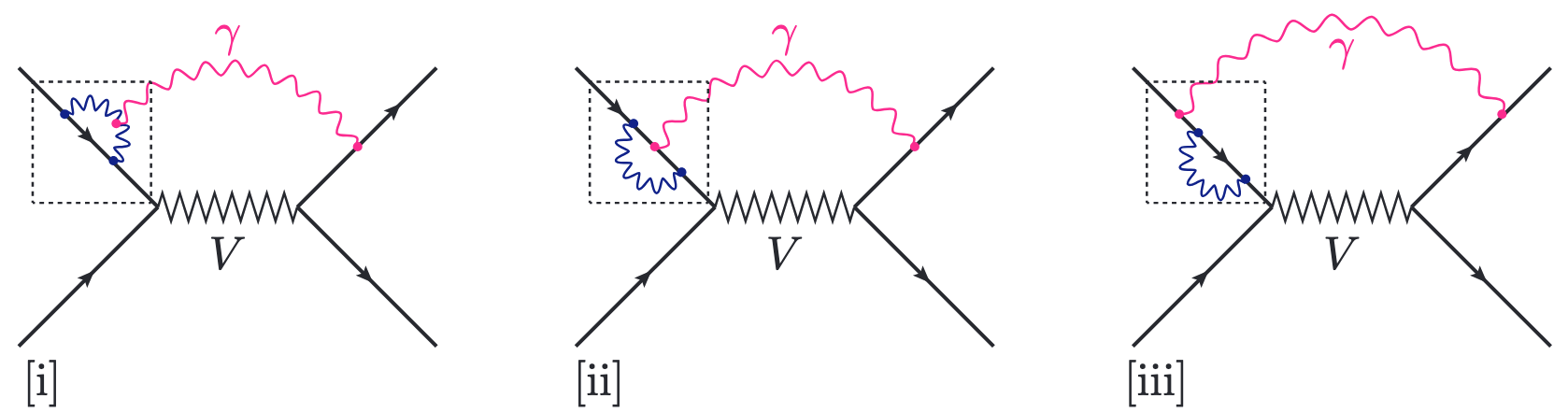}
    \caption{Example of two-loop diagrams involving a bosonic correction to the external fermion line, dressed with a soft photon.}
    \label{fig:FFV}
\end{figure}

\begin{figure}[t]
   \centering
    \includegraphics[width=0.5\textwidth]{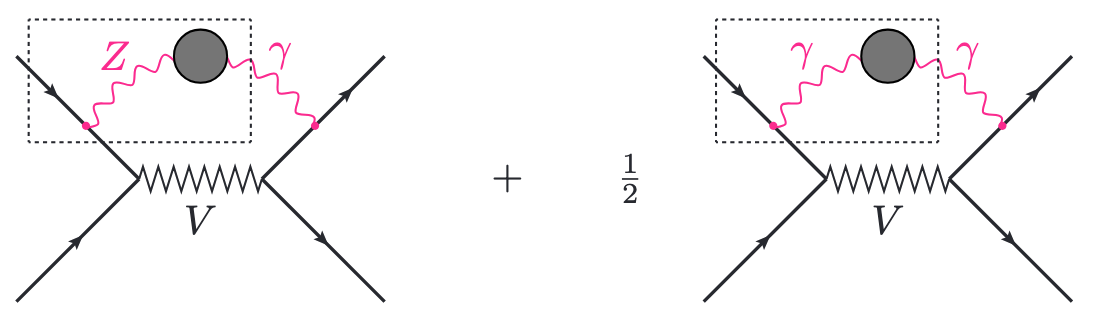}
    \caption{$Z\gamma$-mixing and photon self-energy diagrams.}
    \label{fig:mixing}
\end{figure}

As we are interested in EW corrections, contributions from bosonic loops should also be included.
We begin with the class of diagrams depicted in Figure~\ref{fig:FFV}, where a bosonic correction to an external fermion line is dressed by the exchange of a soft photon with a different fermion line. A similar discussion applies when the soft photon couples to the charged resonance $V$.
 
If the boson in the loop is a $Z$ boson or a photon, the sum of the last two diagrams vanishes identically, independently of the gauge choice.
The situation is more subtle when a $W$ boson is in the loop. At first sight, these diagrams may yield a leading, gauge-dependent contribution that does not fall into any of the three categories (a),(b),(c) discussed in Sec.~\ref{sec:preliminaries}. 
However, a nontrivial cancellation occurs~\cite{Bardin:1999ak} once they are combined with the bosonic contributions to the photon propagator and the $Z\gamma$ mixing (see Figure~\ref{fig:mixing}).  

We give a brief explanation of the origin of this cancellation by considering the subdiagrams highlighted with the dashed lines in Figure~\ref{fig:FFV}. 
Their sum, in the soft-photon limit, stripped of the external spinors, can be recast in the form $\Lambda_V^\mu(0)+\Lambda_{f_j}^\mu(0)$, where
\begin{equation}
    \Lambda_V^{\mu}(0) = \frac{\alpha_0}{4\pi}\,(e_{f_j} \mathcal{V}_1(\epsilon,m_W^2) + \mathcal{V}_2(\epsilon,m_W^2)) (ie \gamma^\mu) \frac{1-\gamma_5}{2}
\end{equation}
is the vertex correction (first two diagrams) split into the $\mathcal{V}_1$ part, proportional to the electric charge $e_{f_j}$ of the external fermion $f_j$, and the remaining part $\mathcal{V}_2$, while
\begin{equation}
    \Lambda_{f_j}^\mu(0)  =  \frac{\alpha_0}{4\pi}\, \Sigma^{(1)}_{f_j}(0) \,e_{f_j} (ie \gamma^\mu) \frac{1-\gamma_5}{2}
\end{equation}    
is the expression of the third diagram. Here, $\Sigma^{(1)}_{f_j}(0)$ is the one-loop bosonic correction to the (massless) fermion self-energy at zero momentum transfer.
Note that the sum of all these contributions is a gauge-dependent quantity. 
However, the gauge dependence is cancelled once the contributions from the highlighted subdiagrams in Figure~\ref{fig:mixing}, namely the $Z\gamma$-mixing term
\begin{equation}
	\Lambda_{ZA}^\mu(0) = \frac{\alpha_0}{4\pi}\, \mathcal{V}_{ZA}(\epsilon,m_W^2) ( ie \gamma^\mu) \left(e_{f_j} \sin^2\theta_W - I_{f_j}^{(3)} \frac{1-\gamma_5}{2}\right)
\label{eq:ZAmixing}
\end{equation}
and the photon self-energy correction
\begin{equation}
	\Lambda_{AA}^\mu(0) =  \frac{\alpha_0}{4\pi}\Pi^{(1),W}_{\gamma\gamma}(0) \,\frac{e_{f_j}}{2} (ie \gamma^\mu)
\label{eq:AAterm}
\end{equation}
are included. In Eq.~\eqref{eq:ZAmixing} $I_{f_j}^{(3)}$ is the eigenvalue of the weak isospin operator, connected to the electric charge $e_{f_j}$ and the weak hypercharge $Y_{f_j}$ by the customary relation $e_{f_j} = I_{f_j}^{(3)} + Y_{f_j}/2$.
The final sum of all diagrams in Figures~\ref{fig:FFV} and \ref{fig:mixing} reads 
\begin{equation}
    \Lambda_V^{\mu}(0)+\Lambda_{f_j}^\mu(0)  + \Lambda_{ZA}^\mu(0)+  \Lambda_{AA}^\mu(0)= 
      \frac{\alpha_0}{4\pi} \mathcal{A}_{\gamma\gamma}^{(1),W}(0) \,\frac{e_{f_j}}{2} (ie \gamma^\mu ) \,,
\label{eq:totalffA}
\end{equation}    
where 
\begin{equation}
    \mathcal{A}_{\gamma\gamma}^{(1),W}(0) 
    = \left(\Pi^{(1),W}_{\gamma\gamma}(0)+ 2\sin^2\theta_W\mathcal{V}_{ZA}(\epsilon,m_W^2)\right)
    \label{eq:gaugeindepA}
\end{equation}
is now a gauge-independent quantity. In the previous expression, $\sin^2\theta_W = 1- m_W^2/m_Z^2$ is the sine of the weak mixing angle.

We observe that, in the unitary gauge, the aforementioned cancellation is manifest: the sum of all diagrams in Figure~\ref{fig:FFV} (i.e. $\Lambda_V^\mu(0) + \Lambda_{f_j}^\mu(0)$) vanishes at leading power, as does the corresponding $Z\gamma$-mixing term $\Lambda_{ZA}^\mu(0)$. 
As a result, the only surviving contribution to the non-factorisable corrections originates from $W$-loop insertions in the soft-photon propagator.  
In contrast, in the Feynman gauge, the expected result is restored only upon summing all contributions, due to non-trivial cancellations between the individual terms on the l.h.s. of \refeq{eq:totalffA}.
Note that, in this example, we considered only the corrections to one of the two external fermions to which the soft photon is attached. 
The same result holds for the other fermion as well.

By bookkeeping all possible contributions~\footnote{There is a caveat when the (bosonic and fermionic) loop correction involves the charged resonance. The cancellation mechanism discussed in the main text holds at the level of the UV renormalised amplitude, as discussed in Sec.~\ref{sec:preliminaries}.} and following the same computational strategy outlined for the massive fermion loops (see Sec.~\ref{sec:fermionic}), we finally obtain
\begin{align}
    \delta_{ab \otimes \gamma\gamma}^{(2),W} = - \delta_{ab}^{(1)} \; \mathcal{A}_{\gamma\gamma}^{(1),W}(0)\,,
    \label{eq:abAAheavyW}
\end{align}
at fixed emitters $a, b \in\{i,f,r\}$ of the soft photon. Here, $\delta_{ab}^{(1)}$ is the corresponding one-loop non-factorisable correction (\refeqs{eq:1lif}–(\ref{eq:1lrr})), and $\mathcal{A}_{\gamma\gamma}^{(1),W}(0)$ is given by
\begingroup
\allowdisplaybreaks
\begin{align}
    \mathcal{A}_{\gamma\gamma}^{(1),W}(0) = -S_\epsilon e^{\gamma_E\epsilon} \Gamma[\epsilon]
    \left[
        \frac{1}{3}(2\epsilon+21)   \left(\frac{m_W^2}{\mu_0^2}\right)^{-\epsilon}
    \right]\,.
    \label{eq:PIheavyW}
\end{align}
\endgroup
We conclude by observing that the remaining bosonic contributions, not explicitly discussed in this section, are either subleading or contributing to $\mathcal{M}_{\text{fact}\otimes \text{nf}}^{(2)}$\,.

\section{Results}
\label{sec:results}

In Sec.~\ref{sec:twoloop}, we discussed the procedure adopted for the computation of the two-loop non-factorisable corrections in a few specific examples.
Here, we will summarise the results for the genuine non-factorisable corrections (Sec.~\ref{sec:results_two-loop-nonfact}), discuss the impact of the UV renormalisation (Sec.~\ref{sec:results_UVren}) and perform the stringent check of infrared pole cancellation (Sec.~\ref{sec:results_IRpoles}). We will conclude our discussion by providing the results for the one- and two-loop finite remainders in PA. 

\subsection{Genuine two-loop non-factorisable corrections}
\label{sec:results_two-loop-nonfact}
The scattering amplitude in \refeq{eq:Mbare} for the fully off-shell process \eqref{eq:process} is obtained as the sum of the two classes of factorisable and non-factorisable corrections.
We first discuss the genuine two-loop non-factorisable corrections involving two soft virtual photons.
By summing over all possible combinations $ab\otimes cd$, where the indices $ab$ ($cd$) denote the pair of edges to which the first (second) photon loop is attached,
and exploiting the results in \refeqs{eq:ifixfx}-(\ref{eq:rrrr}), we obtain 
\begingroup
\allowdisplaybreaks
\begin{align}
    \delta_{\nfbis\otimes \nfbis}^{(2)} &= 
    \frac{\Gamma[1+4\epsilon]}{\Gamma[1+2\epsilon]^2}
    \Bigg[
    \frac{1}{2} \sum_{i, i'  \in \mathcal{I}} \sum_{f, f'  \in \mathcal{F}} \delta_{if}^{(1)}\delta_{i'f'}^{(1)} 
    + \frac{1}{2} \sum_{i, i'  \in \mathcal{I}} \delta_{ir}^{(1)}\delta_{i'r}^{(1)}
    + \frac{1}{2} \sum_{f, f' \in \mathcal{F}} \delta_{rf}^{(1)} \delta_{rf'}^{(1)}
    +\frac{1}{2}\left(\delta_{rr}^{(1)} \right)^2
    + \sum_{i \in \mathcal{I} }\sum_{f \in \mathcal{F} } \delta_{ir}^{(1)} \delta_{rf}^{(1)} 
     \notag \\
    &\hspace{1.5cm}+ \sum_{i, i'  \in \mathcal{I}} \sum_{f \in \mathcal{F}} \delta_{if}^{(1)} \delta_{i'r}^{(1)}
    + \sum_{i \in \mathcal{I} }\sum_{f, f' \in \mathcal{F} } \delta_{if}^{(1)} \delta_{rf'}^{(1)}
    + \delta_{rr}^{(1)} \left( \sum_{i \in \mathcal{I} }  \delta_{ir}^{(1)} + \sum_{f \in \mathcal{F}} \delta_{rf}^{(1)} + \sum_{i \in \mathcal{I} }\sum_{f \in \mathcal{F} } \delta_{if}^{(1)} \right)
    \Bigg]\,,
\end{align}
\endgroup
where $\mathcal{I} (\mathcal{F})$ is the set of IS (FS) charged particles.
This expression is equivalent to
\begingroup
\allowdisplaybreaks
\begin{align}
    \delta_{\nfbis\otimes \nfbis}^{(2)} &=
    \frac{\Gamma[1+4\epsilon]}{2\Gamma[1+2\epsilon]^2}
    \left(\sum_{i\in \mathcal{I}}\sum_{f \in \mathcal{F}} \delta_{if}^{(1)} + \sum_{i\in \mathcal{I}}\delta_{ir}^{(1)} + \sum_{f \in \mathcal{F}} \delta_{rf}^{(1)} + \delta_{rr}^{(1)}\right)^2 
    = \frac{\Gamma[1+4\epsilon]}{2\Gamma[1+2\epsilon]^2} \left(\delta_{\nfbis}^{(1)}\right)^2 \,.
    \label{eq:nf2lfinal}
\end{align}
\endgroup
Therefore, the genuine two-loop non-factorisable corrections arising from purely soft photon exchanges can be written in terms of the one-loop non-factorisable corrections, up to an $\epsilon$-dependent overall normalisation.
Although the individual contributions $\delta_{ab\otimes cd}^{(2)}$ may exhibit $\mathcal{O}(1/\epsilon^4)$ and $\mathcal{O}(1/\epsilon^3)$ poles, as shown in \refeq{eq:ififexp}, the Laurent expansion of the final result in \refeq{eq:nf2lfinal} begins at $\mathcal{O}(1/\epsilon^2)$ .
The cancellation of the higher poles is a direct consequence of charge conservation.
This behaviour is consistent with the soft wide-angle origin of the non-factorisable corrections.

As discussed in Sec.~\ref{sec:twoloop}, the whole two-loop non-factorisable corrections also receive contributions from diagrams in which a single soft photon is exchanged between the production and decay subprocesses, while the second loop corresponds to bosonic or fermionic corrections.
The massless fermionic contributions $\delta_{ab\otimes \gamma\gamma}^{(2),\ell}$, derived in \refeqs{eq:2LifAA}-(\ref{eq:2LrrAA}), lead to
\begingroup
\allowdisplaybreaks
\begin{align}
    \delta_{\nfbis\otimes  \gamma\gamma}^{(2),\ell} &=  S_\epsilon^2 \left(\frac{\mu_0^2}{m_V^2}\right)^{2\epsilon} \left(\frac{-K_V-i0_+}{m_V^2}\right)^{-4\epsilon}\!\! A(\epsilon)  \,\frac{\beta_0}{\epsilon}
    \frac{e^{\gamma_E\epsilon}\Gamma[1+4\epsilon]\Gamma[1-2\epsilon]}{\Gamma[1+\epsilon]}
    \notag\\
    &\times 
     \Bigg[
    \frac{(\sigma_Ve_V)^2}{\epsilon(1-4\epsilon)}
    \!-\frac{\sigma_Ve_V}{4\epsilon^2}\Biggl(\sum_{i\in\mathcal{I}}\sigma_ie_i
    -\sum_{f\in\mathcal{F}}\sigma_fe_f \Biggr)
    + \frac{1}{2\epsilon^2}\sum_{\substack{i\in\mathcal{I}\\f\in\mathcal{F}}}\sigma_ie_i \sigma_fe_f 
    \,{}_2F_1\!\left(1,-2\epsilon;1-2\epsilon;1-\frac{m_V^2}{(2\ONp_i\cdot \ONp_f)}\right)
    \!\Bigg]\,,
\end{align}
\endgroup
upon summing over all pairs of edges $ab$. The above expression can be recast into the following form
\begingroup
\allowdisplaybreaks
\begin{align}
    \delta_{\nfbis \otimes \gamma\gamma}^{(2),\ell} &=
    S_\epsilon^2 \left(\frac{\mu_0^2}{m_V^2}\right)^{2\epsilon}\left(\frac{-K_V-i0_+}{m_V^2}\right)^{\!-4\epsilon} \!\!A(\epsilon)\frac{\beta_0}{\epsilon} \frac{e^{\gamma_E \epsilon} \Gamma[1+4\epsilon] \Gamma[1-2\epsilon]}{\Gamma[1+\epsilon]} \notag \\
    &\times \!\left(\! \frac{\Delta^{(1),-1}_{\nfbis}}{2\epsilon} + \Delta^{(1),0}_{\nfbis} + 2 \epsilon \left( \Delta^{(1),1}_{\nfbis}-\frac{\zeta_2}{2} \Delta^{(1),-1}_{\nfbis}   \right) 
    + 4\epsilon^2\left(\Delta_{\nfbis}^{(1),2} - \frac{\zeta_2}{2} \Delta_{\nfbis}^{(1),0} -\frac{\zeta_3}{3} \Delta_{\nfbis}^{(1),-1}  \right) 
    + \mathcal{O}(\epsilon^3) \!\right) \,,
     \label{eq:nfAAmasslessfinal}
\end{align}
\endgroup
where $\Delta_{\nfbis}^{(1),m}$ is the $m$-th coefficient of the Laurent expansion 
\begingroup
\allowdisplaybreaks
\begin{align}
    \delta_{\nfbis}^{(1)} = S_\epsilon \left(\frac{\mu_0^2}{m_V^2}\right)^{\epsilon}\left(\frac{-K_V-i0_+}{m_V^2}\right)^{-2\epsilon} e^{2 \gamma_E \epsilon} \Gamma[1+2\epsilon]
    \sum_{m=-1}^{\infty} \Delta^{(1),m}_{\nfbis}\epsilon^m
    \label{eq:nf1Lsymb}
\end{align}
\endgroup
of the one-loop non-factorisable corrections~\eqref{eq:deltanf1L}.
The explicit expressions for the first two coefficients,
\begingroup
\allowdisplaybreaks
\begin{align}
	\Delta^{(1),-1}_{\nfbis} &=  2 \sum_{\substack{i\in\mathcal{I}\\f\in\mathcal{F}}}\sigma_ie_i\sigma_fe_f  
\left(\log\left(\frac{m_V^2}{2\ONp_i\cdot \ONp_f}\right)-1\right)  \,,\\
	\Delta^{(1),0}_{\nfbis} &= -2 \sum_{\substack{i\in\mathcal{I}\\f\in\mathcal{F}}}\sigma_ie_i\sigma_fe_f \left( \text{Li}_2\left(1-\frac{m_V^2}{2\ONp_i\cdot \ONp_f}\right) +2  \right)  \,,
      \label{eq:Deltanf1L}
\end{align}
\endgroup
can be straightforwardly obtained from \refeq{eq:deltanf1Lexp}.
On the contrary, the massive contribution is
\begin{align}
    \delta_{\nfbis\otimes\gamma\gamma}^{(2),h} = -\; \delta_{\nfbis}^{(1)} \;  \mathcal{A}_{\gamma\gamma}^{(1),h}(0) \,,
    \label{eq:nfAAheavyfinal}
\end{align}
at the leading power in the expansion parameter $\lambda$, where 
\begingroup
\allowdisplaybreaks
\begin{align}
    \mathcal{A}_{\gamma\gamma}^{(1),h}(0) =  \mathcal{A}_{\gamma\gamma}^{(1),W}(0) +\Pi_{\gamma\gamma}^{(1),F}(0)\,.
    \label{eq:PIheavy}
\end{align}
\endgroup
The explicit expressions of $\Pi_{\gamma\gamma}^{(1),F}(0)$ and $\mathcal{A}_{\gamma\gamma}^{(1),W}(0)$ are given in \refeqs{eq:PIheavyF} and (\ref{eq:PIheavyW}), respectively.

\subsection{UV renormalisation}
\label{sec:results_UVren}

Following the presentation of the final results for the genuine two-loop non-factorisable corrections in Sec.~\ref{sec:results_two-loop-nonfact}, we now address the renormalisation of UV divergences.
Since our focus is on the non-factorisable contributions, we do not discuss in detail the UV renormalisation of external fermion wave-functions and masses in the OS scheme, as these affect only the factorisable corrections, which are kept implicit. 
As noted above, the OS renormalisation of the resonance $V$ is crucial to remove contributions that would otherwise be superleading in the power counting.
After this step, the only remaining impact of UV renormalisation on the two-loop non-factorisable corrections arises from the renormalisation of the electric charge.

We adopt the $\MSbar$ scheme for the charge renormalisation. 
This leads to the following gauge-independent relation,
\begin{align}
   S_\epsilon \alpha_0 \mu_0^{2\epsilon} = \alphaEW\mu^{2\epsilon} \left(1- \frac{1}{\epsilon}\left(\beta_0+\beta_0^h+7\right)\frac{\alphaEW}{4\pi}
    + \mathcal{O}(\alpha^2)\right)  \,,
    \label{eq:chargerenom}
\end{align}
between the bare coupling $\alpha_0$ in \refeq{eq:Mbare} and the UV renormalised EW coupling $\alphaEW$, where light and heavy fermions as well as the $W$ boson contribute to the running.
The first-order contribution to the $\beta$ function due to massive fermions is
\begin{align}
    \beta_0^h = -\frac{4}{3} \sum_{j\in\{h\}} N_c^{j} e_j^2 \,,
\end{align}
while $\beta_0$ is given in \refeq{eq:beta0}.
Having this in mind, the UV renormalised amplitude in the PA is
\begin{align}
    \M_{\PA}(\alphaEW, \mu) &= (4\pi \alphaEW)  \biggl( \M_{\PA}^{(0)}
    + \frac{\alphaEW}{4\pi}\!\left( \Mren_{\fact}^{(1)}(\mu) + \Mren_{\nf}^{(1)}(\mu) \right) \notag \\
    &\hspace{3.25cm}+ \left( \frac{\alphaEW}{4\pi} \right)^{\!\!2}\!
    \left( \Mren_{\fact}^{(2)}(\mu) + \Mren_{\nf}^{(2)}(\mu) \right) + \mathcal{O}(\alpha^3) \biggr)  \,,
    \label{eq:Mren}
\end{align}
where the tilde denotes that both the factorisable and non-factorisable contributions have been renormalised.

As expected, the one-loop non-factorisable corrections
\begin{align}
  \Mren_{\nf}^{(1)} = \deltaren_{\nfbis}^{(1)}\M_{\PA}^{(0)}\,
  \label{eq:nf1Lren}
\end{align}
are not affected by the charge renormalisation~\eqref{eq:chargerenom} and contain purely infrared singularities.
Note that in \refeq{eq:nf1Lren} $\deltaren_{\nfbis}^{(1)}$ corresponds to the result in \refeq{eq:deltanf1L} with the overall normalisation $S_\epsilon$ set to unity and $\mu_0$ replaced by the renormalisation scale $\mu$.

On the contrary, at two-loop order, the UV renormalised non-factorisable corrections are given by
\begingroup
\allowdisplaybreaks
 \begin{align}
   \Mren_{\nf}^{(2)} = 
   \deltaren_{\nfbis}^{(1)} \Mren_{\fact}^{(1)} + \left( \deltaren_{\nfbis\otimes \nfbis}^{(2)} + \deltaren_{\nfbis\otimes \gamma\gamma}^{(2),h}+ \deltaren_{\nfbis\otimes \gamma\gamma}^{(2),\ell} - \frac{1}{\epsilon}\deltaren_{\nfbis}^{(1)} \left(\beta_0  + \beta_0^h + 7 \right)\right)\M_{\PA}^{(0)}\,,
    \label{eq:nf2Lren}
\end{align}
\endgroup
where $\deltaren_{\nfbis\otimes \nfbis}^{(2)}$, $\deltaren_{\nfbis\otimes \gamma\gamma}^{(2),\ell}$ and $\deltaren_{\nfbis\otimes \gamma\gamma}^{(2),h}$ are provided in 
\refeqs{eq:nf2lfinal}, (\ref{eq:nfAAmasslessfinal}), and (\ref{eq:nfAAheavyfinal}), respectively, with $S_\epsilon=1$ and $\mu_0\to\mu$.
The net effect of UV renormalisation on the genuine two-loop non-factorisable corrections, i.e.\ the expression in parentheses, arises solely from the one-loop charge renormalisation in \refeq{eq:chargerenom}.
We observe that, after including the UV mass and wave-function counterterms associated with the resonance $V$, the factorisable corrections dressed with a soft photon ($\M_{\fact\otimes\nfbis}^{(2)}$ introduced in Sec.\ref{sec:preliminaries}) exactly factorise into the product of the one-loop UV renormalised factorisable corrections $\Mren_{\fact}^{(1)}$ and the one-loop non-factorisable coefficient $\deltaren_{\nfbis}^{(1)}$, i.e.\ the first term in \refeq{eq:nf2Lren}.

The last step before discussing the IR structure of the non-factorisable corrections in Sec.~\ref{sec:results_IRpoles}, 
consists of the decoupling of the heavy degrees of freedom.
This can be achieved via a finite renormalisation shift (see e.g. Ref.\cite{Steinhauser:2002rq}) 
\begingroup
\allowdisplaybreaks
\begin{align}
    \alphaEW = \zeta_\alpha  \;\alphaEWnl\,,
    \label{eq:alpha_decoupling}
\end{align}
\endgroup
where $\alphaEWnl$ denotes the coupling in the effective field theory, in which the heavy degrees of freedom coupled to the photon (i.e.\ massive fermions and charged $W$ boson) are integrated out.  
The (gauge-independent) decoupling constant $\zeta_\alpha$ is given by~\cite{Bednyakov:2016onn}
\begingroup
\allowdisplaybreaks
\begin{align}
    \zeta_\alpha = 1 + \frac{\alphaEWnl}{4\pi} \left( \widetilde{\mathcal{A}}_{\gamma\gamma}^{(1),h}(0)  +  \frac{1}{\epsilon} (\beta_0^h + 7)\right) + \mathcal{O}(\alpha^2)
\end{align}
\endgroup
up to one-loop order,
where $\widetilde{\mathcal{A}}_{\gamma\gamma}^{(1),h}(0)$ is the expression in \refeq{eq:PIheavy} with $S_\epsilon=1$ and $\mu_0\to\mu$.
We recall that the non-factorisable corrections are not affected by the decoupling transformation at one-loop order,
while at two loops they are modified as
\begin{align}
\label{eq:final-UVren-nonfact}
   & \Mrendc_{\nf}^{(2)} = \Mren_{\nf}^{(2)} + 2 \zeta_{\alpha}^{(1)} \;\deltaren_{\nfbis}^{(1)}\;\M_{\PA}^{(0)} \,,
\end{align}
where $\zeta_{\alpha}^{(1)} $ denotes the $\mathcal{O}\left({\alpha}\right)$ coefficient of the decoupling constant $\zeta_\alpha$.
We observe that one of the terms proportional to $\zeta_{\alpha}^{(1)}$ in the above equation is absorbed into the definition of the one-loop factorisable corrections in the decoupling scheme,
\begin{align}
       & \Mrendc_{\fact}^{(1)}= \Mren_{\fact}^{(1)} + \zeta_{\alpha}^{(1)} \:\M_{\PA}^{(0)}\,,
\end{align}
which contribute to the first term in \refeq{eq:nf2Lren}. 
The remaining term cancels identically, to all orders in $\epsilon$, against the contributions from heavy degrees of freedom in \refeq{eq:nf2Lren}.
The final expression of the two-loop non-factorisable corrections in the decoupling scheme is 
\begingroup
\allowdisplaybreaks
\begin{align}
    \Mrendc_{\nf}^{(2)} = 
       \deltaren_{\nfbis}^{(1)} \Mrendc_{\fact}^{(1)} + \left( \deltaren_{\nfbis\otimes \nfbis}^{(2)} + \deltaren_{\nfbis\otimes \gamma\gamma}^{(2),\ell} - \frac{\beta_0}{\epsilon}\deltaren_{\nfbis}^{(1)} \right)\M_{\PA}^{(0)}\,.
       \label{eq:nf2Lren_dc-scheme}
\end{align}
\endgroup

\subsection{IR singularity structure and finite remainder}
\label{sec:results_IRpoles}

A stringent check of the correctness of our computation of the non-factorisable corrections is provided by the cancellation of the IR poles. Indeed, the PA preserves the structure of the virtual IR poles in the vicinity of the resonant region.\footnote{Formally, the poles are also preserved outside the resonant region. However, their coefficients are numerically different because they must be evaluated using the on-shell-projected kinematic invariants.}

Writing the UV renormalised EW amplitude in the decoupling scheme~\eqref{eq:alpha_decoupling}
\begin{align}
    \M_{\PA}(\alphaEWnl, \mu) = (4\pi \alphaEWnl) \left(\M_{\PA}^{(0)}
    + \frac{\alphaEWnl}{4\pi} \Mrendc_{\PA}^{(1)}(\mu)
    + \left(\frac{\alphaEWnl}{4\pi}\right)^{\!2} \Mrendc_{\PA}^{(2)}(\mu) + \mathcal{O}(\alpha^3)\right) \,,
\end{align}
with 
\begingroup
\allowdisplaybreaks
\begin{align}
	\Mrendc_{\PA}^{(1)} &= \Mrendc_{\fact}^{(1)} + \Mrendc_{\nf}^{(1)} =  \Mrendc_{\fact}^{(1)} + \deltaren_{\nfbis}^{(1)}\M_{\PA}^{(0)} \,,	\\
	\Mrendc_{\PA}^{(2)} &= \Mrendc_{\fact}^{(2)} + \Mrendc_{\nf}^{(2)} = \Mrendc_{\fact}^{(2)} + \deltaren_{\nfbis}^{(1)} \Mrendc_{\fact}^{(1)} + \left( \deltaren_{\nfbis\otimes \nfbis}^{(2)} + \deltaren_{\nfbis\otimes \gamma\gamma}^{(2),\ell} - \frac{\beta_0}{\epsilon}\deltaren_{\nfbis}^{(1)}  \right)\M_{\PA}^{(0)} \,,
\end{align}
\endgroup
we can predict its IR structure by the subtraction operator \cite{Ferroglia:2009ii,Becher:2009kw,Becher:2009qa,Becher:2009cu}
\begin{equation}
    \mathcal{Z}(\mu) = 1 + \frac{\alphaEWnl}{4 \pi} \mathcal{Z}^{(1)}(\mu) + \left(\frac{\alphaEWnl}{4 \pi}\right)^2 \mathcal{Z}^{(2)}(\mu) + \mathcal{O}(\alpha^3)\,,
    \label{eq:Neubert_Ioperator}
\end{equation}
where 
\begingroup
\allowdisplaybreaks
\begin{align}
 \Z^{(1)} &= \frac{\Gamma_0'}{4 \epsilon^2} + \frac{\Gamma_0}{2 \epsilon}   \label{eq:Neubert_Ioperator_1L} \,, \\
 \Z^{(2)} &= \frac{\Gamma_0^{'} \cdot \Gamma_0^{'}}{32 \epsilon^4} + \frac{1}{8 \epsilon^3}\left(\Gamma_0^{'} \cdot \Gamma_0^{}-\frac{3}{2}\beta_0 \Gamma_0^{'}\right) + \frac{1}{8 \epsilon^2}\left(\Gamma_0^{} \cdot \Gamma_0^{}-2\beta_0 \Gamma_0^{}\right) + \frac{\Gamma_1^{'}}{16 \epsilon^2} + \frac{\Gamma_1^{}}{4 \epsilon}  \label{eq:Neubert_Ioperator_2L}\,.
\end{align}
\endgroup
The anomalous dimensions appearing in the previous equation, which control the pole structure, are obtained from their QCD counterparts through an \textit{abelianisation} procedure. Their explicit expressions are provided in Appendix~\ref{app:2}.
For the reader's convenience, in \refeqs{eq:anomdimFULL1}-(\ref{eq:anomdimFULL2}) we also give their expression in the specific case of a massless four-fermion scattering.

Therefore, the IR poles are expected to take the following form up to two-loop order:
\begingroup
\allowdisplaybreaks
\begin{align}
    \Mrendc^{(1)}_{\PA}\vert_{\sing} &= \Z^{(1)} \M_{\PA}^{(0)} \,, \label{eq:sing1l} \\
    \Mrendc^{(2)}_{\PA}\vert_{\sing} &= \left[ \Z^{(2)} - \left(\Z^{(1)}\right)^2 \right] \M_{\PA}^{(0)} + \left( \Z^{(1)} \Mrendc_{\PA}^{(1)} \right)\Bigl|_{\mathrm{poles}} 
    = \Z^{(2)} \M_{\PA}^{(0)} + \left(\Z^{(1)} \Mrendc_{\PA}^{(1), \text{fin}}\right)\Bigl|_{\mathrm{poles}}  \,, \label{eq:sing2l}
\end{align}
\endgroup
where the $d$-dimensional Born $\M_{\PA}^{(0)}$ is always factored out, and we have exploited the fact that
\begin{equation}
    \Mrendc_{\PA}^{(1)} = \Z^{(1)} \M_{\PA}^{(0)} + \Mrendc_{\PA}^{(1), \text{fin}}
\end{equation}
with $\Mrendc_{\PA}^{(1), \text{fin}}$ regular for $\epsilon \to 0$.\footnote{Note that higher orders in $\epsilon$ in the one-loop amplitude have to be taken into account.}
The singularity structure presented in \refeqs{eq:sing1l} and (\ref{eq:sing2l}) also holds for the purely factorisable amplitudes, provided that the subtraction operator $\Z$~\eqref{eq:Neubert_Ioperator} is replaced by its fully factorised counterpart $\Z_{\fact}$.
The expressions for the one- and two-loop coefficients of $\Z_{\fact}$ are analogous to those in \refeqs{eq:Neubert_Ioperator_1L}-(\ref{eq:Neubert_Ioperator_2L}), upon replacing the anomalous dimensions $\Gamma^{'}$ and $\Gamma$ with $\Gamma^{'}_{\fact}$ and $\Gamma_{\fact}$, respectively.
The definition of the factorised anomalous dimensions controlling the IR poles of the amplitudes associated with the on-shell production and decay subprocesses is provided in Appendix~\ref{app:2}.

Consequently, the pole structure of the non-factorisable corrections at $n$-loop order can be predicted by taking the difference between the IR singularities of the PA and factorisable amplitudes, i.e.\
\begin{align}
    \Mrendc_{\nf}^{(n)}\vert_{\sing} = \Mrendc_{\PA}^{(n)}\vert_{\sing} - \Mrendc_{\fact}^{(n)}\vert_{\sing}\,.
\end{align}
At one-loop order, this translates into
\begingroup
\allowdisplaybreaks
\begin{align}
    \Mrendc_{\PA}^{(1)}\vert_{\sing} - \Mrendc_{\fact}^{(1)}\vert_{\sing} =\left( \Z^{(1)} - \Z_{\fact}^{(1)}\right)\M^{(0)}_{\PA}
    =  \frac{2}{\epsilon}\biggl[ \sum_{\substack{i\in \mathcal{I}\\f\in\mathcal{F}}} \sigma_ie_i\sigma_f e_f \log\left(\frac{m_V^2}{2\ONp_i\cdot \ONp_f}\right) + e_V^2
    \biggr] \M^{(0)}_{\PA}\,,
\end{align}
\endgroup
where the expected poles coincide with those found in \refeq{eq:deltanf1Lexp} from the explicit computation of the one-loop non-factorisable corrections.

Similarly, the expected poles of the two-loop non-factorisable corrections can be extracted from the difference
\begingroup
\allowdisplaybreaks
\begin{align}
    \Mrendc_{\PA}^{(2)}\vert_{\sing} - \Mrendc_{\fact}^{(2)}\vert_{\sing} &=
    \left(\Z^{(2)} -\Z_{\fact}^{(2)} \right)\M_{\PA}^{(0)} 
    + \left(\Z^{(1)}\Mrendc_{\PA}^{(1),\fin}-\Z_{\fact}^{(1)}\Mrendc_{\fact}^{(1),\fin}\right)\Bigl|_{\mathrm{poles}}  \,,
    \label{eq:diff2l}
\end{align}
\endgroup
which can be recast into the form
\begingroup
\allowdisplaybreaks
\begin{align}
    \Mrendc_{\PA}^{(2)}\vert_{\sing} - \Mrendc_{\fact}^{(2)}\vert_{\sing} &=
    \left(\Z^{(2)} -\Z_{\fact}^{(2)} \right)\M_{\PA}^{(0)} 
    + \left[ \deltaren_{\nfbis}^{(1)}\vert_{\sing} \Mrendc_{\fact}^{(1),\fin}
    + \left( \Z_{\fact}^{(1)} 
    + \deltaren_{\nfbis}^{(1)}\vert_{\sing} \right)\deltaren_{\nfbis}^{(1),\fin} \M_{\PA}^{(0)}  \right]\Bigl|_{\mathrm{poles}} \,,
    \label{eq:diff2Lsing}
\end{align}
\endgroup
since 
\begin{align}
    \Mrendc_{\PA}^{(1)} = \left(\Z^{(1)}_{\fact} + \deltaren_{\nfbis}^{(1)}\vert_{\sing}\right) \M_{\PA}^{(0)}
     + \left(\Mrendc_{\fact}^{(1),\fin}+\deltaren_{\nfbis}^{(1),\fin}\M_{\PA}^{(0)}\right)\,.
\end{align}
At the same time, the IR singularities of the UV renormalised two-loop non-factorisable amplitude can be extracted from its explicit expression~\eqref{eq:nf2Lren_dc-scheme} by retaining only the terms that diverge in the $\epsilon\to 0$ limit, i.e.\
\begingroup
\allowdisplaybreaks
\begin{align}
    \Mrendc_{\nf}^{(2)}\vert_{\sing} &=
   \left[ \deltaren_{\nfbis}^{(1)}\vert_{\sing} \Mrendc_{\fact}^{(1),\fin}
    + \left( \Z_{\fact}^{(1)} 
    + \deltaren_{\nfbis}^{(1)}\vert_{\sing} \right)\deltaren_{\nfbis}^{(1),\fin} \M_{\PA}^{(0)}  \right]\Bigl|_{\mathrm{poles}} 
     \notag\\
    &
    + \left[ \deltaren_{\nfbis}^{(1)}\vert_{\sing} \Z^{(1)}_{\fact} 
    +\frac{1}{2} \left(\deltaren_{\nfbis}^{(1)}\vert_{\sing} \right)^2
    + \left( \deltaren_{\nfbis\otimes \gamma\gamma}^{(2),\ell}  -\frac{\beta_0}{\epsilon} \deltaren_{\nfbis}^{(1)} \right)\Bigl|_{\mathrm{poles}}  \right] \M_{\PA}^{(0)} 
    \label{eq:nf2Lsing}\,.
\end{align}
\endgroup
We note that the terms in the first line are identical to those appearing in the square brackets of \refeq{eq:diff2Lsing}.
This observation allows us to verify the cancellation of the poles without requiring the knowledge of the higher-order terms in $\epsilon$ of the one-loop amplitude in PA.

Since the $1/\epsilon^4$ poles cancel exactly in the difference $(\Z^{(2)} -\Z_{\fact}^{(2)})$ in \refeq{eq:diff2Lsing}, the leading divergence in \refeq{eq:nf2Lsing} starts at $\mathcal{O}(\epsilon^{-3})$.
It arises from the product of the $1/\epsilon$ soft wide-angle singularity of the one-loop non-factorisable coefficient $\deltaren^{(1)}_{\nfbis}$ and the $1/\epsilon^2$ soft-collinear pole of the one-loop factorisable corrections.
We explicitly checked that all $\epsilon$ poles cancel between the second line of \refeq{eq:nf2Lsing} and the difference $(\Z^{(2)} -\Z_{\fact}^{(2)})$ in \refeq{eq:diff2Lsing}.

Having discussed the subtraction of the IR poles, we can now define the finite remainder in the decoupling $\MSbar$ scheme as 
\begin{align}
	\mathbb{F}_{\PA}(\alphaEWnl) = \left(\mathcal{Z}^{-1}(\mu)\,\M_{\PA}(\alphaEWnl,\mu)\right)\Big|_{\epsilon\to 0} = \mathbb{F}_{\fact} + \mathbb{F}_{\nf}  \,.
\label{eq:IRfinrem}
\end{align}
Therefore, we can extract the one- and two-loop coefficients of the $\mathbb{F}_{\nf}$ in the usual $\frac{\alphaEWnl}{4\pi}$ expansion as
\begingroup
\allowdisplaybreaks
\begin{align}
\label{eq:fin1nonfact}
    \mathbb{F}_{\nf}^{(1)} = & \left(\Delta_{\nfbis}^{(1),0} - 2 \log\left(\frac{-K_V}{\mu\, m_V}\right)\Delta_{\nfbis}^{(1),-1}\right)\M_{\PA}^{(0)}
    \\
    \label{eq:fin2nonfact}
    \mathbb{F}_{\nf}^{(2)} = & \Delta_{\nfbis}^{(1),0} \mathbb{F}_{\fact}^{(1)} 
    +  \left[\frac{1}{2}\left(\Delta_{\nfbis}^{(1),0}\right)^2 + 2 \zeta_2 \left(\Delta_{\nfbis}^{(1),-1}\right)^2
    +\beta_0 \left(\frac{5}{3} \Delta_{\nfbis}^{(1),0} + \left(\frac{14}{9}+\frac{3}{2} \zeta_2 \right)\Delta_{\nfbis}^{(1),-1} + \Delta_{\nfbis}^{(1),1}\right)\right]\M^{(0)}_{\PA}
    \notag\\
    &- 2 \log\left(\frac{-K_V}{\mu\, m_V}\right)\left(\Delta_{\nfbis}^{(1),-1 } \mathbb{F}_{\fact}^{(1)}  + \left[\Delta_{\nfbis}^{(1),0 }\Delta_{\nfbis}^{(1),-1}+\beta_0 \left(\Delta^{(1),0}_{\nfbis} + \frac{5}{3}\Delta^{(1),-1}_{\nfbis} \right)\right]\M^{(0)}_{\PA}
    \right)
    \notag\\
    &+ 2 \log^2\left(\frac{-K_V}{\mu\, m_V}\right)\Delta_{\nfbis}^{(1),-1} \left(\Delta_{\nfbis}^{(1),-1}+\beta_0 \right)\M^{(0)}_{\PA}\,,
\end{align}
\endgroup
where the definition of $\Delta_{\nfbis}^{(1),m}$ follows from \refeq{eq:nf1Lsymb}.

\section{Discussion}
\label{sec:discussion}

The results of Sec.~\ref{sec:results} show that, after UV renormalisation and proper decoupling of heavy degrees of freedom, the two-loop non-factorisable corrections can be written (see \refeq{eq:final-UVren-nonfact}) as the sum of three contributions, supplemented by a UV counterterm.
The first term, $\deltaren_{\nfbis}^{(1)} \Mrendc_{\fact}^{(1)}$, corresponds to the product of the one-loop factorisable and non-factorisable corrections. The second, $\deltaren_{\nfbis\otimes \nfbis}^{(2)}$, is obtained through an iteration of the one-loop result (see \refeq{eq:nf2lfinal}). The third contribution, $\deltaren_{\nfbis\otimes \gamma\gamma}^{(2),\ell}$, arises from the insertion of a light-fermion loop in the soft-photon propagator.

It is natural to conjecture that the iterated structure of \refeq{eq:nf2lfinal} can be generalised to $n$-loop order as
\begin{align}
    \delta^{(n)}_{\nfbis\otimes \nfbis} &=  \frac{\Gamma[1+2n\epsilon]}{n!}\left(\frac{\delta_{\nfbis}^{(1)}}{\Gamma[1+2\epsilon]}\right)^n\,,
    \label{eq:nloop}
\end{align}
a relation that we have explicitly verified up to three loops ($n=3$). This structure is expected on general grounds as a consequence of the fact that multiple soft-photon emissions are uncorrelated \cite{Yennie:1961ad}.
We can therefore anticipate that at least a part of the non-factorisable corrections can be resummed to all orders.
More precisely, within the PA, the amplitude is expected to factorise into a {\it hard} contribution---corresponding to the factorisable corrections--and a {\it soft} contribution, in which large logarithms of the ratio between a soft scale, $\Gamma_V$, and the hard scale, $m_V$, are resummed to all orders.

Before presenting our ansatz for a possible all-order resummation, we point out that the iterated structure suggested by \refeq{eq:nloop} is not exact. Indeed, at each perturbative order, a factor $\Gamma[1+2n\epsilon]/\Gamma[1+2\epsilon]^n$ appears that spoils the exact factorisation picture.
The origin of this factor can be traced back to the shift in the power of the resonance propagator induced by the loop integrations (see e.g.\ \refeq{eq:deltaifif2}), which leads to $\M_\nf^{(1)} \sim K_V^{-1-2\epsilon}$, $\M_{\nfbis \otimes \nfbis}^{(2)} \sim K_V^{-1-4\epsilon}$, and so forth. Indeed, at $n$-loop order, a coupled resonant denominator of the form $(2(l_1 + l_2 + \dots + l_n)\cdot p_V + K_V)^{-1}$ appears at the integrand level, as seen explicitly for $n=2$. Such a coupled multi-loop propagator can be disentangled by a Laplace transform
\begin{equation}
	\frac{1}{2(l_1 + l_2+\dots +l_n)\cdot p_{V}+K_V} = \mathcal{L}\left[\prod_{i=1}^n e^{-2l_i\cdot p_{V} t}\right] \,,
\end{equation}
where $\mathcal{L}$ denotes the Laplace transform of a function $f(t)$ with $t$ being the conjugate variable of the resonance (almost on-shell) propagator $K_V$.
This observation suggests defining a dimensionful variable
\begin{equation}
	\Delta \equiv (-K_V - i 0_+)/m_V
\end{equation}
and its Laplace conjugate $t$, which may be interpreted as the formation time of the resonance. 
In the conjugate space, we can write down an all-order resummation formula for the finite remainder in PA (see \refeq{eq:IRfinrem}) as
\begin{equation}
	\mathcal{L}^{-1}[\mathbb{F}_{\PA}(\alpha(m_V))] = \mathcal{H}(\alpha(m_V)) \,\text{exp}\left\{ -\frac{1}{2}\int_{\musoftsq}^{m_V^2} \frac{d\mu^2}{\mu^2} \Gamma_s(\alpha(\mu)) \right\} \, \mathcal{S}(\alpha(\musoft)) \,, \quad \bar{t} \equiv t e^{\gamma_{E}}
	\label{eq:factorisation}
  \end{equation}
which resums large logarithmic corrections of the form $\ln(\bar{t}\,m_V)$ in the conjugate space, or equivalently $\ln(m_V/\Delta)$ in momentum space. 
In \refeq{eq:factorisation}, up to next-to-leading logarithmic (NLL) accuracy, the hard function 
\begingroup 
\allowdisplaybreaks
\begin{equation}
	\mathcal{H}(\alpha(m_V)) = (4 \pi \alpha(m_V))\left( \mathcal{A}^{(0)} +\frac{\alpha(m_V)}{4\pi}\mathcal{A}_\fact^{(1)}(m_V) + \left(\frac{\alpha(m_V)}{4\pi}\right)^2\mathcal{A}_\fact^{(2)}(m_V) +\mathcal{O}(\alpha^3) \right)
	\label{eq:hard_function}
\end{equation}
\endgroup
collects the contributions from the purely factorisable corrections (hard modes), while the soft function 
\begingroup
\allowdisplaybreaks
\begin{equation}
	\mathcal{S}(\alpha(\musoft)) = e^{\varSigma_s(\alpha(\musoft))} = 1 + \frac{\alpha(\musoft)}{4\pi}\mathcal{S}^{(1)} + \left(\frac{\alpha(\musoft)}{4\pi}\right)^{\!2}\mathcal{S}^{(2)} +\mathcal{O}(\alpha^3)
	\label{eq:soft_function}
\end{equation}
\endgroup
corresponds to the boundary term of the non-factorisable corrections at the soft scale $\musoft$ (soft modes).
Note that $\mathcal{A}^{(0)}$ and $\mathcal{A}_\fact^{(n)}$ in \refeq{eq:hard_function} denote the Born $\M_{\PA}^{(0)}$ and $n$-loop finite remainder of the factorisable corrections $\mathbb{F}_{\fact}^{(n)}$, respectively, with the factor $1/\Delta$ stripped.

The soft function in \refeq{eq:soft_function} is expected to exponentiate, with
\begingroup
\allowdisplaybreaks
\begin{equation}
	\varSigma_s(\alpha(\musoft)) = \frac{\alpha(\musoft)}{4\pi}\varSigma_s^{(1)} + \left(\frac{\alpha(\musoft)}{4\pi}\right)^{\!2} \varSigma_s^{(2)} +\mathcal{O}(\alpha^3) \,,
	\label{eq:soft_function_exponent}
\end{equation}
\endgroup
encoding the genuinely new contributions occurring order by order in perturbation theory.
The explicit expression of the one- and two-loop coefficients of the soft function is
\begin{align}
	\mathcal{S}^{(1)} &= \Delta_{\nfbis}^{(1),0} \,,\\
	\mathcal{S}^{(2)} &= \frac{1}{2}\left(\Delta_{\nfbis}^{(1),0}\right)^2 -\beta_0 \Delta_{\nfbis}^{(1),1} + \beta_0 \left(\left(\frac{14}{9} - \frac{\zeta_2}{2}\right)\Delta_{\nfbis}^{(1),-1} + \frac{5}{3} \Delta_{\nfbis}^{(1),0} + 2 \Delta_{\nfbis}^{(1),1}  \right) \,,
\end{align}
or analogously
\begin{align}
	\varSigma_s^{(1)} &= \Delta_{\nfbis}^{(1),0} \,,\\
	\varSigma_s^{(2)} &= \beta_0 \left(\left(\frac{14}{9} - \frac{\zeta_2}{2}\right)\Delta_{\nfbis}^{(1),-1} + \frac{5}{3} \Delta_{\nfbis}^{(1),0} + \Delta_{\nfbis}^{(1),1}  \right) \,,
\end{align}
where the $m$-th coefficient $\Delta_{\nfbis}^{(1),m}$ of the Laurent expansion in \refeq{eq:nf1Lsymb} can be extracted from the all-order one-loop result in \refeq{eq:deltanf1L}.

The soft anomalous dimension in \refeq{eq:factorisation} admits a perturbative expansion in the coupling, 
\begin{equation}
	\Gamma_s(\alpha(\mu)) = \frac{\alpha(\mu)}{4\pi}\Gamma_s^{(1)} + \left(\frac{\alpha(\mu)}{4\pi}\right)^{\!2}\Gamma_s^{(2)} +\mathcal{O}(\alpha^3) \,,
	\label{eq:soft_anomalous_dimension}
\end{equation}
and controls the RG evolution of the soft function from $\musoft$ to the hard scale $m_V$. 
Each coefficient $\Gamma_s^{(n)}$ can be extracted from the $1/\epsilon$ poles (multiplied by a factor $-2n$) of the genuinely new (i.e.\ not obtained from iterations of lower-order kernels) UV renormalised $n$-loop non-factorisable corrections (after having set $\bar{t}=1$).
Therefore, up to two loops, the anomalous-dimension coefficients are
\begin{align}
  \Gamma_s^{(1)} = -2 \Delta_{\nfbis}^{(1),-1} ~,~~\quad ~~~ \Gamma_s^{(2)} =2 \kappa \Gamma_s^{(1)}~,~~\quad ~~~ \kappa = - \frac{10}{9}\sum_{i \in \{\ell\}} N_c^i e_i^2 \,,
\label{eq:kappa}  
\end{align}
where $\kappa$ is the abelianised version of the coefficient $K = \left(\frac{67}{18} - \zeta_2 \right) C_A - \frac{10}{9}T_R N_f$, with $T_R = 1/2$ and $N_f$ the number of active quark flavours.
It follows that \refeq{eq:soft_anomalous_dimension} up to NLL accuracy is entirely and universally controlled by the use of an effective soft-photon coupling 
\begin{align}
	\alpha^{\mathrm{eff}}(\mu) = \alphaEWnl \left( 1+ \frac{\alphaEWnl}{2 \pi}\kappa \right) \,,
\end{align}
which corresponds to the abelianisation of the strong coupling in the Catani-Marchesini-Webber (CMW) scheme~\cite{Catani:1990rr}.
Equivalently, the coefficient $\kappa$ in \refeq{eq:kappa} is just the abelian version of the second-order coefficient \cite{Curci:1980uw,Furmanski:1980cm} of the cusp anomalous dimension~\cite{Brandt:1981kf,Korchemsky:1987wg} (see also \refeq{eq:massless_cusp} in Appendix \ref{app:2}).

When expanded to ${\cal O}(\alpha^2)$ and transformed back in $\Delta$ space, the resummation formula in \refeq{eq:factorisation} exactly reproduces the results of Sec.~\ref{sec:results}. In particular it predicts the correct $\log^2(\Delta)$ and $\log(\Delta)$ terms in Eqs.~(\ref{eq:fin1nonfact}-\ref{eq:fin2nonfact}).
Although we have not carried out a complete calculation, we expect our
all-order picture to be valid also at three-loop order.
In particular, since the contribution of three photons attached to a
closed fermion loop vanishes by Furry's theorem, we expect the third-order
soft anomalous dimension $\Gamma_{s}^{(3)}$ to be governed by the abelianisation of the cusp anomalous dimension at the same order. We leave a detailed investigation of this to future work. Starting
from four loops, we expect genuinely new quartic contributions
arising from massless fermion loops analogous to those considered for single
soft emission in Ref.~\cite{Ma:2023gir}. The four or more virtual photons can
attach to the initial- and final-state charged particles and to the resonance
itself in all possible ways. We foresee these diagrams to be leading in the
power counting, thereby entering the factorisation picture at
higher perturbative orders.

Finally, we comment on the extension of our results to processes involving multiple resonances.
We expect that the factorisation picture no longer holds at the integrated level due to a non-trivial dependence of the loop integrals on the momenta of multiple resonances. 
More investigations are left for future work.

\section{Conclusions}
\label{sec:conclusions}


In this paper, we have studied virtual EW corrections to $2 \to 2$ fermion scattering processes mediated by a vector boson, $V = W^\pm, Z$, within the PA. In this framework, two classes of gauge-invariant contributions are identified.
The {\it factorisable} corrections comprise all resonant terms that can be expressed as the product of on-shell matrix elements describing the production and decay subprocesses, connected by the kinematics and spin of the intermediate resonance. Their evaluation requires the knowledge of on-shell, polarised amplitudes for production and decay, which are simpler to compute as they involve fewer external legs and scales than the full off-shell process.
The {\it non-factorisable} corrections account for the remaining resonant contributions and arise from soft-photon exchanges between the initial- and final-state charged particles and/or the resonance itself.

We have focused on the non-factorisable contributions and performed an explicit two-loop calculation, expressing the result in terms of suitable soft integrals. After evaluating these integrals, we showed that, once the effects of charged heavy degrees of freedom (i.e.\ heavy fermions and the $W$ boson) are properly decoupled, the two-loop corrections can be written in dimensional regularisation as an iteration of the corresponding one-loop contribution, supplemented by a genuinely new term induced by light-fermion loops.
The resulting two-loop non-factorisable corrections exhibit poles up to $1/\epsilon^2$, as expected from their soft, wide-angle origin. Upon combining their infrared singularities with those from the factorisable contributions, the expected singularity structure of the full amplitude is correctly reproduced.

The factorisable corrections, while not yet available in a form directly suitable for our purposes, are in principle known in the literature (see Ref.~\cite{Chen:2022dow} and references therein).
Combined with the non-factorisable results presented here, they can be used to construct an approximate expression for the two-loop EW amplitude for key scattering processes such as Drell-Yan lepton-pair production in hadron collisions and $e^+e^- \to \mu^+\mu^-$. Such a prediction within the PA may also serve as a useful benchmark for future complete computations of the two-loop amplitudes for these processes.

Finally, we have outlined how our analysis can be extended beyond fixed order. 
In particular, we have proposed an all-order resummation formula that captures the factorisation of the amplitude into a hard contribution---corresponding to the factorisable corrections---and a soft component, in which large logarithms of the ratio between a soft scale, $\Gamma_V$, and the hard scale, $m_V$, are resummed to all orders. This structure is consistent with our explicit results up to two-loop order.
Given the soft origin of the non-factorisable corrections, the insights gained in this work are expected to provide useful guidance also in the more involved QCD case, where the soft quanta are self-interacting gluons.

The relatively simple factorised picture emerging from our computation, however, is specific to single-resonance production. In processes involving multiple resonances, the soft momentum does not flow uniquely between the initial and final states, thus leading to a more intricate structure of the loop integrals. We leave a detailed investigation of these cases to future work.

\vspace*{0.7cm}
\noindent {\bf Acknowledgements}    

\noindent 
We would like to thank Thomas Gehrmann, Giampiero Passarino, Stefano Pozzorini and Adrian Signer for discussions.
We also thank Jonas Lindert and Alessandro Vicini for comments on the manuscript.
This work is supported in part by the Swiss National Science Foundation (SNF) under contract 200020$\_$219367 and by the Swiss High Energy Physics initiative for the FCC (CHEF), with funding provided specifically by SERI and the University of Zurich.
The work of C.S. is partly supported by the Excellence Cluster ORIGINS, funded
by the Deutsche Forschungsgemeinschaft (DFG, German Research Foundation) under Germany’s
Excellence Strategy — EXC-2094-390783311. The work of L.B. is funded by the
European Union (ERC, grant agreement No. 101044599, JANUS). Views and opinions expressed
are however those of the authors only and do not necessarily reflect those of the European Union or
the European Research Council Executive Agency. Neither the European Union nor the granting
authority can be held responsible for them.

\appendix

\section{Expressions of the scalar one- and two-loop eikonal integrals}
\label{app:1}

In this appendix, we list the analytic results for the eikonal integrals that appear in the computation of the two-loop non-factorisable corrections in Sec.~\ref{sec:two_soft-photons} and Sec.~\ref{sec:fermionic}. 
Since in the two-loop calculation one-loop integrals appear with shifted propagator powers, we provide, for completeness, the expressions for the one-loop box ($\D$), triangle ($\C$), and bubble ($\B$) integral families.

We denote with $\mathcal{J}[\{r,j_1,\dots, j_n\};a_0,a_1]$ a generic one-loop integral family with $n+1$ linear propagators. 
The indices $j_1,\dots,j_n \in \mathcal{I} \cup \mathcal{F}$ stand for the massless external momenta $p_{j_1},\dots,p_{j_n}$ appearing in the linear propagators $1/(2l\cdot p_j + i0_+)$. The index $r$ labels the resonance momentum $p_r$ in the integral denominator $(2l\cdot p_r + K_r)$, where $K_r = p_r^2 - \mu_r^2$ is the off-shell propagator of the resonance $r$. 
The real parameters $a_0$ and $a_1$ denote the powers of the quadratic propagator $1/l^2$ and the resonance propagator $1/(2l\cdot p_r + K_r)$, respectively.
Having this notation in mind, the expressions of the required one-loop integral families are given by
\begingroup
\allowdisplaybreaks
\begin{align}
     \D[\{r,i,f\};a_0,a_1]
    =&\, \mu_0^{4-d}\int \frac{d^dl}{(2\pi)^d}\frac{1}{(l^2)^{a_0}(2l\cdot p_r+K_r)^{a_1}(2l\cdot p_i+i0_+)(2l\cdot p_f+i0_+)} 
    \notag\\
    =& -e^{i\pi (a_0+a_1)} 
    \frac{i \;S_\epsilon }{(4\pi)^{2}}    \left(\frac{-K_r-i0_+}{p_r^2}\right)^{\!-2\epsilon} \!\left(\frac{\mu_0^2}{p_r^2}\right)^{\!\epsilon}
    \frac{e^{\gamma_E\epsilon}\Gamma[1-a_0-\epsilon]\Gamma[2a_0+a_1-2+2\epsilon]}{(a_0-1+\epsilon)\Gamma[a_0]\Gamma[a_1]} 
    \notag\\
    &\times(-K_r-i0_+)^{-2a_0-a_1+2}\frac{(p_r^2)^{a_0-1}}{2p_i\cdot p_f}{}_2F_1\!\left(\!1,-a_0+1-\epsilon;-a_0+2-\epsilon;1-\frac{(2p_i\cdot p_r)(2p_r\cdot p_f)}{p_r^2 \;(2p_i\cdot p_f)}\right)  
    \label{eq:D}
\end{align}
\begin{align}
     \C[\{r,i\};a_0,a_1]
    =&\, \mu_0^{4-d}\int \frac{d^dl}{(2\pi)^d}\frac{1}{(l^2)^{a_0}(2l\cdot p_r+K_r)^{a_1}(2l\cdot p_i+i0_+)} 
    \notag\\
    =& -e^{i\pi (a_0+a_1)} 
    \frac{i\;S_\epsilon}{(4\pi)^{2}}\left(\frac{-K_r-i0_+}{p_r^2}\right)^{\!-2\epsilon} \!\left(\frac{\mu_0^2}{p_r^2}\right)^{\!\epsilon}
    \frac{e^{\gamma_E\epsilon}\Gamma[1-a_0-\epsilon]\Gamma[2a_0+a_1-3+2\epsilon]}{\Gamma[a_0]\Gamma[a_1]} 
   \notag\\
   &\times (-K_r-i0_+)^{-2a_0-a_1+3}\frac{(p_r^2)^{a_0-1}}{2p_i\cdot p_r}
    \label{eq:C}  
\end{align}
\begin{align}
    \B[\{r\};a_0,a_1]
    =&\, \mu_0^{4-d}\int \frac{d^dl}{(2\pi)^d}\frac{1}{(l^2)^{a_0}(2l\cdot p_r+K_r)^{a_1}} 
    \notag\\
    =&\, e^{i\pi (a_0+a_1)} 
    \frac{i\;S_\epsilon}{(4\pi)^{2}}     \left(\frac{-K_r-i0_+}{p_r^2}\right)^{\!-2\epsilon} \!\left(\frac{\mu_0^2}{p_r^2}\right)^{\!\epsilon}
    \frac{e^{\gamma_E\epsilon}\Gamma[2-a_0-\epsilon]\Gamma[2a_0+a_1-4+2\epsilon]}{\Gamma[a_0]\Gamma[a_1]} 
    \notag\\
    &\times(-K_r-i0_+)^{-2a_0-a_1+4}(p_r^2)^{a_0-2}
    \label{eq:B}
\end{align}
\endgroup
where $\mu_0$ is the dimensional-regularisation scale and $S_\epsilon = (4\pi)^\epsilon e^{-\gamma_E\epsilon}$.

For the two-loop eikonal integrals, we introduce the generic notation $\mathcal{J}\!\mathcal{Y}[\{L_1\},\{L_2\},r]$. 
The set $\{L_1\}$ ($\{L_2\}$) specifies the indices of the external massless momenta $p_j$ ($j \in \mathcal{I} \cup \mathcal{F}$) and/or resonance momentum $p_r$ appearing in the linear propagators that depend only on the loop momentum $l_1$ ($l_2$). Each entry of the set follows the same convention described for the one-loop integrals. 
The quadratic propagators $l_1^2$ and $l_2^2$ are implicitly understood in this notation. 
The additional index $r$ not belonging to the lists $\{L_1\}, \{L_2\}$ denotes the resonance momentum appearing in the coupled integral denominator $(2l_1\cdot p_r + 2l_2\cdot p_r + K_r)$.
Following this notation, the expressions of the required two-loop soft integrals are
\begingroup
\allowdisplaybreaks
\begin{align}
    \DD[\{i,f\},\{i',f'\},r] =& 
    \int \frac{d^dl_1d^dl_2}{(2\pi)^{2d}}\frac{(\mu_0^2)^{4-d} \,(2l_1\cdot p_{r}+2l_2\cdot p_{r}+K_r)^{-1}}{l_1^2l_2^2(2l_1\cdot p_i+i0_+)(2l_1\cdot p_f+i0_+)(2l_2\cdot p_{i'}+i0_+)(2l_2\cdot p_{f'}+i0_+) }
    \notag\\
    =& -\frac{S_\epsilon^2}{(4\pi)^4}     
    \left(\frac{-K_r-i0_+}{p_r^2}\right)^{-4\epsilon}\left(\frac{\mu_0^2}{p_r^2}\right)^{2\epsilon}
    \frac{e^{2\gamma_E\epsilon}\Gamma[1-\epsilon]^2\Gamma[1+4\epsilon]}{\epsilon^4}
    \frac{1}{K_r(2p_i\cdot p_f)(2p_{i'}\cdot p_{f'})}
    \notag\\
    &\times{}_2F_1\left(1,-\epsilon;1-\epsilon;1-\frac{(2p_i\cdot p_r)(2p_r\cdot p_f)}{p_r^2 \;(2p_i\cdot p_f)}\right)
   {}_2F_1\left(1,-\epsilon;1-\epsilon;1-\frac{(2p_{i'}\cdot p_r)(2p_r\cdot p_{f'})}{p_r^2 \;(2p_{i'}\cdot p_{f'})}\right)
\end{align}
\begin{align}
    \DC[\{i,f,r\},\{i'\},r] =&  \int \frac{d^dl_1d^dl_2}{(2\pi)^{2d}}\frac{(\mu_0^2)^{4-d}\, (2l_1\cdot p_r+2l_2\cdot p_r +K_r)^{-1}}{l_1^2l_2^2(2l_1\cdot p_i+i0_+)(2l_1\cdot p_f+i0_+)(2l_1\cdot p_r+K_r)(2l_2\cdot p_{i'}+i0_+)}
    \notag\\
    =&\, \frac{S_\epsilon^2}{(4\pi)^4}
    \left(\frac{-K_r-i0_+}{p_r^2}\right)^{-4\epsilon}\left(\frac{\mu_0^2}{p_r^2}\right)^{2\epsilon}
    \frac{e^{2\gamma_E\epsilon}\Gamma[1-\epsilon]^2 \Gamma[1+4\epsilon]}{2\epsilon^4}
    \frac{1}{K_r(2p_i\cdot p_{f})(2p_r\cdot p_{i'})}
    \notag\\
    &\times{}_2F_1\left(1,-\epsilon;1-\epsilon;1-\frac{(2p_i\cdot p_r)(2p_r\cdot p_f)}{p_r^2 \;(2p_i\cdot p_f)}\right)
\end{align}
\begin{align}
      \DB[\{i,f,r\},\{\},r;a_0] =&\int\frac{d^dl_1d^dl_2}{(2\pi)^{2d}}
   \frac{(\mu_0^2)^{4-d} }{l_1^2 l_2^2(2l_1\cdot p_i+i0_+)(2l_1\cdot p_f+i0_+)(2l_1\cdot p_r+K_r)^{a_0} (2l_1\cdot p_r+2l_2\cdot p_r +K_r)}
\notag\\
=&\,- e^{i\pi a_0} \frac{S_\epsilon^2}{(4\pi)^4} 
    \left(\frac{-K_r-i0_+}{p_r^2}\right)^{-4\epsilon}\left(\frac{\mu_0^2}{p_r^2}\right)^{2\epsilon}
\frac{e^{2\gamma_E\epsilon}\Gamma[1-\epsilon]^2\Gamma[1+2\epsilon]\Gamma[a_0-1+4\epsilon]}{2\epsilon^3(1-2\epsilon)\; \Gamma[a_0-1+2\epsilon]}
   \notag\\
    &\times\frac{(-K_r-i0_+)^{-a_0+1}}{p_r^2(2p_i\cdot p_f)}
    {}_2F_1\left(1,-\epsilon;1-\epsilon;1-\frac{(2p_i\cdot p_r)(2p_r\cdot p_f)}{p_r^2 \;(2p_i\cdot p_f)}\right)
\end{align}
\begin{align}
    \CC[\{i,r\},\{f\},r] =&\,  \int \frac{d^dl_1d^dl_2}{(2\pi)^{2d}}\frac{(\mu_0^2)^{4-d}}{l_1^2 l_2^2(2l_1\cdot p_i+i0_+)(2l_1\cdot p_r+K_r)(2l_2\cdot p_f+i0_+)(2l_1\cdot p_r+2l_2\cdot p_r + K_r)}
     \notag\\
     =&\, -\frac{S_\epsilon^2}{(4\pi)^4}
         \left(\frac{-K_r-i0_+}{p_r^2}\right)^{\!-4\epsilon}\!\left(\frac{\mu_0^2}{p_r^2}\right)^{\!2\epsilon}
     \frac{e^{2\gamma_E\epsilon}\Gamma[1-\epsilon]^2\Gamma[1+4\epsilon]}{8\epsilon^4}
     \frac{1}{(2p_r\cdot p_i)(2p_r\cdot p_f)}
\end{align}
\begin{align}
    \CB[\{i,r\},\{\},r;a_0] =&\, \int \frac{d^dl_1d^dl_2}{(2\pi)^{2d}}\frac{(\mu_0^2)^{4-d}}{l_1^2 l_2^2(2l_1\cdot p_i+i0_+)(2l_1\cdot p_r+K_r)^{a_0}(2l_1\cdot p_r+2l_2\cdot p_r + K_r)}
    \notag\\
    =& - e^{i\pi a_0}\frac{S_\epsilon^2}{(4\pi)^4}
        \left(\frac{-K_r-i0_+}{p_r^2}\right)^{\!-4\epsilon}
        \!
        \left(\frac{\mu_0^2}{p_r^2}\right)^{2\epsilon} \notag \\
     &\times \frac{e^{2\gamma_E\epsilon}\Gamma[1-\epsilon]^2\Gamma[1+2\epsilon]\Gamma[a_0-2+4\epsilon]}{2\epsilon^2(1-2\epsilon)\;\Gamma[a_0-1+2\epsilon]} 
     \frac{(-K_r-i0_+)^{-a_0+2}}{p_r^2(2p_r\cdot p_f)}
\end{align}
\begin{align}
    \CB[\{i\},\{r\},r] &= \int \frac{d^dl_1d^dl_2}{(2\pi)^{2d}}
    \frac{ (\mu_0^2)^{4-d}}{l_1^2 l_2^2(2l_1\cdot p_i+i0_+)(2l_2\cdot p_r+K_r)(2l_1\cdot p_r+2l_2\cdot p_r + K_r)}
    \notag\\
    &= -\frac{S_\epsilon^2}{(4\pi)^4} 
    \left(\frac{-K_r-i0_+}{p_r^2}\right)^{-4\epsilon}\left(\frac{\mu_0^2}{p_r^2}\right)^{2\epsilon}
    \frac{e^{2\gamma_E\epsilon}\Gamma[1-\epsilon]^2\Gamma[1+4\epsilon]}{8\epsilon^3(1-4\epsilon)}
    \frac{K_r}{p_r^2(2p_r\cdot p_f)}
\end{align}
\begin{align}
    \BB[\{r\},\{\},r; a_0] =&\, \int\frac{d^dl_1d^dl_2}{(2\pi)^{2d}} \frac{(\mu_0^2)^{4-d}}{l_1^2l_2^2(2l_1\cdot p_r+K_r)^{a_0}(2l_1\cdot p_r+2l_2\cdot p_r+K_r)}
    \notag\\
    =& -e^{-i\pi a_0}\frac{S_\epsilon^2}{(4\pi)^4} 
    \left(\frac{-K_r-i0_+}{p_r^2}\right)^{-4\epsilon}\left(\frac{\mu_0^2}{p_r^2}\right)^{2\epsilon} \notag \\
    &\times \frac{e^{2\gamma_E\epsilon}\Gamma[1-\epsilon]^2 \Gamma[1+2\epsilon]\Gamma[a_0-3+4\epsilon]}{2\epsilon(1-2\epsilon)\Gamma[a_0-1+2\epsilon]}
    \frac{ (-K_r-i0_+)^{-a_0+3} }{(p_r^2)^2}\,.
\end{align}
\endgroup
In the above equations, $\DB$, $\CB$, and $\BB$ take an additional argument $a_0$, which specifies the power of the propagator $1/(2l_1\cdot p_r+K_r)$.

The expressions of the integrals presented in this appendix have been numerically cross-checked against \texttt{AMFlow}~\cite{Liu:2022chg} for a couple of benchmark points.

\section{Expressions of the anomalous dimensions}
\label{app:2}
As mentioned earlier, the expression of the anomalous dimensions appearing in Eq.~\eqref{eq:Neubert_Ioperator} can be extracted from the QCD counterpart~\cite{Ferroglia:2009ii,Becher:2009kw,Becher:2009qa,Becher:2009cu} via an \textit{abelianisation} procedure. 
This procedure involves replacing gluons with photons and applying the appropriate replacement rules for the colour factors.
Therefore, the expressions of the QED anomalous dimensions are given by
\begingroup
\allowdisplaybreaks
\begin{align}
    \Gamma' &= \sum_{n=0}^{\infty} \left(\frac{\alphaEWnl}{4\pi}\right)^{n+1} \Gamma_n' = -2 \sum_j e_j^2 \gamma_{\text{cusp}}(\alpha) \,, 
    \label{eq:anomalous_dimensions1}
    \\
    \Gamma &= \sum_{n=0}^{\infty} \left(\frac{\alphaEWnl}{4\pi}\right)^{n+1} \Gamma_n = 
    \sum_{j\ne j'} \frac{(\sigma_j e_j)(\sigma_{j'} e_{j'})}{2}
     \gamma_{\text{cusp}}(\alpha)\log\left( \frac{\mu^2}{-s_{jj'}} \right) + \sum_j \gamma^j(\alpha)
     \notag \\ 
     &- \sum_{J \ne J'} \frac{(\sigma_J e_J)(\sigma_{J'} e_{J'})}{2} \gamma_{\text{cusp}}(\beta_{JJ'},\alpha) 
    + \sum_J \gamma^J(\alpha) + \sum_{J,j} (\sigma_j e_j) (\sigma_J e_J)\gamma_{\text{cusp}}(\alpha) \log\left(\frac{m_J\mu}{-s_{Jj}}\right) \,,
    \label{eq:anomalous_dimensions}
\end{align}
\endgroup
where the anomalous-dimension coefficients are defined as
\begingroup
\allowdisplaybreaks
\begin{align}
    \gamma^j(\alpha) &= \frac{\alphaEWnl}{4\pi}(-3 e_j^2) + \left(\frac{\alphaEWnl}{4\pi} \right)^2 \left[ e_j^4 \left( -\frac{3}{2}+2\pi^2 -24\zeta_3\right) - \frac{\beta_0}{2}  \,e_j^2 \left(\frac{65}{9} + \pi^2 \right) \right] + \mathcal{O}(\alpha^3) \,,
    \\
    \gamma^J(\alpha) &= \frac{\alphaEWnl}{4\pi}(-2 e_J^2) + \left(\frac{\alphaEWnl}{4\pi} \right)^2 \left(-\frac{10}{3}\beta_0\,e_J^2\right) + \mathcal{O}(\alpha^3) \,,
\end{align}
and 
\begin{align}
    &\gamma_{\text{cusp}}(\alpha) = \frac{\alphaEWnl}{4 \pi}4 + \left(\frac{\alphaEWnl}{4 \pi}\right)^2 \frac{20}{3}\beta_0 + \mathcal{O}(\alpha^3) \,,
    \label{eq:massless_cusp}
    \\
    &\gamma_{\text{cusp}}(\beta_{JJ'},\alpha) = \gamma_{\text{cusp}}(\alpha) \beta_{JJ'}\coth\beta_{JJ'}\,.
    \label{eq:massive_cusp}
\end{align}
\endgroup
We use lower-case Latin indices $j,j'$ to denote massless fermions (both quarks and leptons in the set $\{ \ell \}$), and upper-case indices $J,J'$ to denote massive charged particles (both fermions and gauge bosons).
The velocity-dependent cusp anomalous dimension is defined in terms of the cusp angle $\beta_{JJ'} = \cosh^{-1}\left(\frac{-s_{JJ'}}{2 m_J m_{J'}}\right)$, while the scalar products $s_{\alpha\beta} = 2 \sigma_{\alpha\beta} \ONp_{\alpha} \cdot \ONp_{\beta}$
are functions of the external momenta $\{\ONp_\kappa\}$ entering the scattering process.
Here $\sigma_{\alpha\beta} = +1$ if both particles are incoming or outgoing, $\sigma_{\alpha\beta} = -1$ otherwise.

We now specialise the general expressions above to the scattering process defined in \refeq{eq:process}. 
As all external legs correspond to massless fermions, terms involving upper-case indices $J$ do not contribute, and the anomalous dimensions reduce to
\begingroup
\allowdisplaybreaks
\begin{align}
    \Gamma &= \sum_{\substack{j\neq j'\\j,j'\in  \mathcal{I}\cup \mathcal{F}}}\frac{(\sigma_j e_j)(\sigma_{j'} e_{j'})}{2}  
       \gamma_{\text{cusp}}(\alpha)\log\left( \frac{\mu^2}{-s_{jj'}} \right) + \sum_{j\in \mathcal{I}\cup \mathcal{F}} \gamma^j(\alpha) \,,
    \label{eq:anomdimFULL1}\\
     \Gamma' &= -2\sum_{j\in \mathcal{I}\cup \mathcal{F}} e_j^2 \gamma_{\text{cusp}}(\alpha) \,.
     \label{eq:anomdimFULL2}
\end{align}
\endgroup
As discussed in Section~\ref{sec:results_IRpoles}, to predict the IR structure of the non-factorisable corrections, the fully factorised version of the subtraction operator is needed.
This requires the anomalous dimensions for the production and decay subprocesses defined in \refeq{eq:subprocess}.
They can be obtained from \refeqs{eq:anomalous_dimensions1}-(\ref{eq:anomalous_dimensions}) by restricting the sums over massless fermionic indices to the subsets $\mathcal{I}$ and $\mathcal{F}$, respectively.
In contrast to \refeq{eq:anomdimFULL1}, contributions from a massive particle $J$ have to be included if the resonance is electrically charged.
Therefore, the corresponding expressions using the on-shell kinematics i.e. $ (\ONp_1+\ONp_2)^2= (\ONp_3+\ONp_4)^2 = m_V^2$ read
\begingroup
\allowdisplaybreaks
\begin{align}
    \Gamma_{\produc} &= -(e_{f_1}e_{f_2})  
       \gamma_{\text{cusp}}(\alpha)\log\left( \frac{\mu^2}{-2\ONp_1\cdot \ONp_2} \right) + \sum_{j\in \mathcal{I}} \gamma^j(\alpha)
        + \gamma^V(\alpha) - e_V^2 \gamma_{\text{cusp}}(\alpha)\log\left( \frac{\mu^2}{m_V^2} \right)  \,,
       \\
    \Gamma'_{\produc} &= -2\sum_{j\in \mathcal{I}} e_j^2 \gamma_{\text{cusp}}(\alpha)  \,,
\end{align}
and
\begin{align}
    \Gamma_{\dec} &= -(e_{f_3}e_{f_4})
       \gamma_{\text{cusp}}(\alpha)\log\left( \frac{\mu^2}{-2\ONp_3\cdot \ONp_4} \right) + \sum_{j\in \mathcal{F}} \gamma^j(\alpha)
       + \gamma^V(\alpha) -  e_V^2 \gamma_{\text{cusp}}(\alpha)\log\left( \frac{\mu^2}{m_V^2} \right) \,,
    \\
     \Gamma'_{\dec} &= -2\sum_{j\in \mathcal{F}} e_j^2 \gamma_{\text{cusp}}(\alpha) \,.
\end{align}
\endgroup
Using the expressions derived above, we can now construct the factorised version of the anomalous dimension as
\begin{align}
    \Gamma_{\fact} &= \Gamma_{\produc} \otimes 1 + 1\otimes \Gamma_{\dec}\, \label{eq:anomdimFACT1}
    \\
    \Gamma'_{\fact} &= \Gamma'_{\produc} \otimes 1 + 1\otimes \Gamma'_{\dec} \label{eq:anomdimFACT2} 
\end{align} 
where $\otimes$ denotes the spin correlations between the production and decay amplitudes.

\bibliography{biblio}

\end{document}